%% file: ms.tex
\documentclass[11pt,twocolumn,tighten,resetfootnote]{aastex701}
\usepackage[utf8]{inputenc}
\usepackage{amsmath}
\usepackage{graphicx}

\usepackage{amssymb}
\usepackage{booktabs}
\usepackage{array}
\usepackage{multirow}
\let\tablenum\relax 
\usepackage{siunitx}

\makeatletter
\@ifpackagelater{array}{2024/01/01}{%
  \ifx\@mkpream@relax\@undefined\else
    \def\aas@startpbox#1{\unexpanded\expandafter{\expandafter\@startpbox\expandafter{#1}}}%
    \g@addto@macro\@mkpream@relax{%
      \let\ar@align@mcell\relax
      \let\do@row@strut\relax
      \let\@endpbox\relax
      \let\@startpbox\aas@startpbox
    }%
  \fi
}{}
\makeatother

\newcommand{\ipac}{Caltech/IPAC, Pasadena, CA 91125, USA}
\newcommand{\carnegie}{Carnegie Science Observatories, Pasadena, CA 91101, USA}

\newcommand{\epl}{Carnegie Science Earth and Planets Laboratory, Washington DC 20015, USA}

\newcommand{\cfa}{Center for Astrophysics $\vert$ Harvard \& Smithsonian, Cambridge, MA 02138, USA}

\newcommand{\komaba}{Komaba Institute for Science, The University of Tokyo, 3-8-1 Komaba, Meguro, Tokyo 153-8902, Japan}
\newcommand{\abc}{Astrobiology Center, 2-21-1 Osawa, Mitaka, Tokyo 181-8588, Japan}

\newcommand{\utokyograd}{Department of Multi-Disciplinary Sciences, Graduate School of Arts and Sciences, The University of Tokyo, 3-8-1 Komaba, Meguro, Tokyo 153-8902, Japan}
\newcommand{\iac}{Instituto de Astrof\'isica de Canarias (IAC), 38205 La Laguna, Tenerife, Spain}

\newcommand{\unc}{Department of Physics and Astronomy, The University of North Carolina at Chapel Hill, Chapel Hill, NC 27599, USA}
\newcommand{\lascampanas}{Las Campanas Observatory, Carnegie Institution for Science, Casilla 601, La Serena, Chile}
\newcommand{\union}{Department of Physics and Astronomy, Union College, 807 Union Street, Schenectady, NY 12308, USA}
\newcommand{\ames}{NASA Ames Research Center, Moffett Field, CA 94035, USA}
\newcommand{\baeri}{Bay Area Environmental Research Institute, Moffett Field, CA 94035, USA}
\newcommand{\ucsc}{Department of Astronomy and Astrophysics, University of California, Santa Cruz, 95064, USA}
\newcommand{\princeton}{Department of Astrophysical Sciences, Princeton University, 4 Ivy Lane, Princeton, NJ 08544, USA}

\input{vals}

\shortauthors{Kathuria et al.}
\shorttitle{An Infant Sub-Saturn in Scorpius-Centaurus}

\begin{document}

\title{An Infant Sub-Saturn in Scorpius-Centaurus\footnote{This paper includes data gathered with the 6.5 meter Magellan Telescopes located at Las Campanas Observatory, Chile.}}

\author[0009-0008-8051-0831]{Gurmeher~Kathuria}
\affiliation{University of California, Los Angeles, CA 90024, USA}
\email{gurmehersk@g.ucla.edu}

\author[0000-0002-0514-5538]{Luke~G.~Bouma}
\affiliation{\ipac}
\affiliation{\carnegie}
\email{lbouma@ipac.caltech.edu}

\author[0000-0001-6037-2971]{Andrew~W.~Boyle}
\altaffiliation{NSF Graduate Research Fellow}
\affiliation{\unc}
\email{awboyle@unc.edu}

\author[0000-0001-6708-3427]{Roberto~Tejada~Arevalo}
\affiliation{\ucsc}
\affiliation{\princeton}
\email{roaareva@ucsc.edu}

\author[0000-0003-3654-1602]{Andrew~W.~Mann}
\affiliation{\unc}
\email{awmann@unc.edu}

\author[0000-0003-2535-3091]{Nidia~Morrell}
\affiliation{\lascampanas}
\email{nmorrell@carnegiescience.edu}

\author[0000-0001-6588-9574]{Karen~A.~Collins}
\affiliation{\cfa}
\email{karen.collins@cfa.harvard.edu}

\author[0000-0003-1305-3761]{R.~Paul~Butler}
\affiliation{\epl}
\email{bluaper@gmail.com}

\author[0000-0002-8681-6136]{Stephen~A.~Shectman}
\affiliation{\carnegie}
\email{shec@carnegiescience.edu}

\author[0009-0008-2801-5040]{Johanna~Teske}
\affiliation{\epl}
\affiliation{\carnegie}
\email{jteske@carnegiescience.edu}

\author[0000-0002-5226-787X]{Jeffrey~D.~Crane}
\affiliation{\carnegie}
\email{crane@carnegiescience.edu}

\author[0000-0003-2527-1475]{Shreyas~Vissapragada}
\affiliation{\carnegie}
\email{svissapragada@carnegiescience.edu}

\author[0000-0002-1715-1257]{Madeleine~McKenzie}
\affiliation{\carnegie}
\email{madeleine.l.mckenzie@gmail.com}

\author[0000-0003-1464-9276]{Khalid~Barkaoui}
\affiliation{\iac}
\affiliation{Astrobiology Research Unit, Universit\'e de Li\`ege, 19C All\'ee du 6 Ao\^ut, 4000 Li\`ege, Belgium}
\affiliation{Department of Earth, Atmospheric and Planetary Science, Massachusetts Institute of Technology, 77 Massachusetts Avenue, Cambridge, MA 02139, USA}
\email{khalid.barkaoui@uliege.be}

\author[0000-0001-8511-2981]{Norio~Narita}
\affiliation{\komaba}
\affiliation{\abc}
\affiliation{\iac}
\email{narita@g.ecc.u-tokyo.ac.jp}

\author[0000-0002-4909-5763]{Akihiko~Fukui}
\affiliation{\komaba}
\affiliation{\iac}
\email{afukui@g.ecc.u-tokyo.ac.jp}

\author[0000-0002-6424-3410]{Jerome~P.~de~Leon}
\affiliation{\komaba}
\email{jpdeleon@g.ecc.u-tokyo.ac.jp}

\author[0009-0007-2926-1924]{Toshi~Suganuma}
\affiliation{\utokyograd}
\email{suganuma-toshi548@g.ecc.u-tokyo.ac.jp}

\author[0000-0002-8399-472X]{Madyson~G.~Barber}
\altaffiliation{NSF Graduate Research Fellow}
\affiliation{\unc}
\email{madysonb@live.unc.edu}

\author[0000-0003-2127-8952]{Francis~P.~Wilkin}
\affiliation{\union}
\email{wilkinf@union.edu}

\author[0009-0001-4431-2909]{Victoria~Zaledonis}
\affiliation{\union}
\email{torizale@icloud.com}

\author{Sophia~Garcia-Gallet}
\affiliation{\union}
\email{garciags@union.edu}

\author[0009-0008-1272-9081]{Glauk~Hizmo}
\affiliation{\union}
\email{glauk.hizmo@icloud.com}

\author[0000-0002-2532-2853]{Steve~B.~Howell}
\affiliation{\ames}
\email{steve.b.howell@nasa.gov}

\author[0000-0002-5741-3047]{David~R.~Ciardi}
\affiliation{\ipac}
\email{ciardi@ipac.caltech.edu}

\author[0000-0001-7746-5795]{Colin~Littlefield}
\affiliation{\baeri}
\email{littlefield@baeri.org}

\author[0009-0002-9833-0667]{Sarah~Deveny}
\affiliation{\ames}
\affiliation{\baeri}
\email{deveny@baeri.org}

\author[0000-0003-1963-9616]{Douglas~A.~Caldwell}
\affiliation{SETI Institute, Mountain View, CA 94043, USA}
\email{dcaldwell@seti.org}

\begin{abstract}
We report the discovery and validation of TIC~88297141\,Ab, a
transiting planet orbiting a $0.31\,M_\odot$, $0.74\,R_\odot$
pre-main-sequence M dwarf in the Scorpius-Centaurus OB association.
Group membership implies an age of $15.6 \pm 1.6$\,Myr,
supported by lithium and isochrone age-dating.  We
detected the TESS transits using a custom pipeline designed for
variable stars and used ground-based photometry (LCO/Swope and
LCOGT/Sinistro) to confirm the on-target nature of the signal.  Our
global fit of the transits and stellar activity gives an orbital
period of 4.64 days and a planet radius of $7.21\pm0.32$\,R$_\oplus$
($0.64\,R_{\rm J}$).   The star has a comoving $0.39\,R_\odot$ M-dwarf
companion at a projected separation of 260\,AU;
speckle imaging (Gemini-S/Zorro) reveals no additional stellar
companions, and radial velocities from Magellan/PFS are consistent
with stellar activity, giving $M_{p}\sin i < 4.7\,M_{\rm J}$ at
$2\sigma$.  A statistical validation analysis yields a false
positive probability of $\fpp$, validating the planet.  A mass
measurement will determine whether TIC~88297141\,Ab is a low-density
super-puff, like other comparably sized planets at these ages, or 
an unusually massive planet for its radius.
\end{abstract}


\section{Introduction}
\label{sec:intro}

A planet's youth is its most dynamic phase, yet its least observed.
Within the first tens of millions of years after formation, planets
are expected to contract, migrate, and shed atmospheres at rates that
exceed anything they will experience over the remainder of their
lives \citep{lin1996, marley2007, owenwu2013}. These same processes can 
also destabilize the resonant chains of adolescent multi-planet systems, 
triggering significant orbital upheaval \citep{izidoro2017}. For a young
sub-Saturn (5--10\,$R_\oplus$), two of these processes are
expected to dominate.  Newly formed planets with substantial gaseous
envelopes are larger and more luminous than their older counterparts,
contracting via Kelvin--Helmholtz cooling on a timescale of roughly
10--100 Myr \citep{marley2007, linder2019}.  Atmospheric mass loss,
whether driven by stellar high-energy irradiation \citep{owenwu2013}
or by the cooling luminosity of the planet's own core
\citep{ginzburg2018}, can meanwhile strip a significant fraction of
the envelope over the following $\sim$10 Myr to 1 Gyr.  Yet the
overwhelming majority of known exoplanets orbit stars with ages
$\gtrsim 1$ Gyr: the Kepler stellar sample has a median age of $\sim
4.6$ Gyr \citep{berger2020}.  Most of what we know about planets is
therefore inferred from systems that have already completed their most
dynamic phase.

Testing these predictions is difficult because young transiting
planets are rare, for two compounding reasons.  Nearby young stars are
intrinsically uncommon: given the Milky Way's roughly constant star
formation rate over its $\sim$10--13\,Gyr history, only a small
fraction of stars in the solar neighborhood are younger than a few
tens of Myr at any given time \citep{binney2000}.  Young stars are
also photometrically hostile targets: starspot modulation and
frequent flaring dominate their light curves, obscuring shallow
transit signals and generating transit-like false positives that
confound pipelines tuned for older, quiescent stars.

An efficient route around the first obstacle is to search for transits
around stars whose youth is already established.  Membership in a
cluster or OB association remains the best-calibrated age indicator
available, since coeval groups can be dated as a whole rather than
star by star \citep{soderblom2010,soderblom2014,krumholz2019}.
Spatial and kinematic clustering from Gaia provides the baseline for
defining the target lists for nearby young populations
\citep{Kerr2021,Hunt2023,Gagne2026}.  Indicators intrinsic to
individual stars, such as photospheric lithium \citep{jeffries2023}
and rotation \citep{Bouma2023}, can then corroborate any given group's
age.  NASA's K2 mission first demonstrated this approach at scale,
surveying Upper Scorpius, Praesepe, and the Hyades along the ecliptic
and returning the first transiting planets younger than 20 Myr,
including K2-33b \citep{david2016, mann2016a}.  The Transiting
Exoplanet Survey Satellite \citep[TESS;][]{ricker2015} extended this
effort to the full sky, and in particular to nearby young stars
amenable to detailed follow-up, with examples including AU~Mic
\citep{plavchan2020,Zicher2022,Donati2025}, TOI-837
\citep{bouma2020,barragan2024,Mantovan2026}, and DS~Tuc
\citep{newton2019b,Zhou2020,Montet2020}.  

The richest hunting ground for the youngest such planets is
Scorpius-Centaurus (Sco-Cen), the nearest OB association with recent
high-mass star formation \citep{dezeeuw1999}, whose $\sim$\,1--20 Myr
population largely traces back to a dominant burst of star formation
roughly 15 Myr ago \citep{Ratzenbock2023b}.  Sco-Cen's youth and
proximity have already yielded planets across detection methods.
Direct imaging has recovered PDS 70 b, still embedded in the disk gap
of its host star \citep{keppler2018}, the two-planet system around
TYC 8998-760-1 \citep{bohn2020a}, and HD 143811 AB b, a $5.6 \pm
1.1\,M_{\rm J}$ companion to a spectroscopic binary \citep{jones2025}.
The transiting census comprises the Jovian-radius HIP~67522\,b and its
near-resonant sibling HIP~67522\,c
\citep{rizzuto2020, barber2024a}, the sub-Jovian TOI-1227\,b \citep{mann2022}, the
super-Neptune TIC~88785435\,b, \citep{vach2025}, and the long-period
giants HD~114082\,b and c \citep{zakhozhay2022, delburgo2026},
together spanning ages of $\sim$11--17\,Myr.  Early results from the
still-small population of infant transiting planets are informative:
atmosphere-based mass constraints from JWST and HST for HIP~67522\,b
and V1298~Tau\,b and c \citep{thao2024b, barat2024a, barat2024b},
together with transit-timing derived masses for the V1298 Tau system
\citep{livingston2026}, suggest that young Jovian-radius planets are
typically super-Earths to Neptunes in mass
($\approx$3--20\,M$_\oplus$) rather than true gas giants. Occurrence-rate
studies agree with this picture, finding that short-period planets are 
more common, and larger at younger ages, with the radius distribution 
steepening over time as planets contract and lose mass \citep{Fernandes2025, Vach2024}.
Yet only about ten transiting planets larger than $\sim$5\,R$_\oplus$ are known below
20\,Myr, and most lack the mass and atmospheric constraints needed to further
test this picture. 

Here, we present the validation and characterization of
TIC~88297141\,Ab, a $\sim$7.2\,R$_\oplus$ sub-Saturn transiting the
primary component of a resolved binary, with a bound stellar companion
at a projected separation of $\sim$260\,AU.  The host star is a
kinematic member of the Sco-Cen EOM~9 subgroup \citep{Kerr2021},
implying an isochrone-derived age of $15.6 \pm 1.6$\,Myr; an
alternative assignment to the $\eta$ Lup subgroup
\citep{Ratzenbock2023a} is discussed and disfavored in
Section~\ref{sec:membership}.  We find no evidence for additional
transiting planets in the system.  TIC~88297141\,Ab occupies a size
regime and age that remain sparsely sampled even as the broader
Sco-Cen census grows. 

The remainder of this paper is organized as follows.
Section~\ref{sec:obs} describes our TESS photometry and ground-based
follow-up. Section~\ref{sec:star} characterizes the system's
properties and assesses its Sco-Cen membership.
Section~\ref{sec:transit} presents our global modeling of the TESS and
ground-based photometry. Section~\ref{sec:validation} evaluates
false-positive scenarios and derives a false-positive probability for
TIC~88297141\,Ab. Finally, Section~\ref{sec:discussion} places the
planet in the context of known young systems and describes
opportunities for follow-up.

\section{Observations}
\label{sec:obs}

\subsection{TESS Photometry}
\label{sec:tess}

Target selection is a crucial problem in studies of stellar
associations, particularly those that are diffuse like Sco-Cen.  We
compiled 12393 candidate Sco-Cen members with $T<14$ from literature
studies based on Gaia positions, velocities, and photometry
\citep{damiani2019, kounkel2020, zerjal2023, Ratzenbock2023b,
Kerr2023, morales2023}, supplemented with high-confidence
eROSITA/eRASS1 sources with optical counterparts in Sco-Cen
\citep{schmitt2022}.  TESS observed 5294 of these stars during Cycle
7, an initial installment of a larger legacy survey; 1985 were
observed for the first time, including TIC~88297141.  We searched for
transiting planet candidates in the Sectors 87--94 Pre-search Data
Conditioning Simple Aperture Photometry (PDCSAP) \citep{Stumpe2012, Stumpe2014, Smith2012} light curves produced
by the Science Processing Operations Center (SPOC) pipeline
\citep{Jenkins2016}. 

\begin{figure*}[p]
    \centering
    \includegraphics[width=0.95\textwidth]{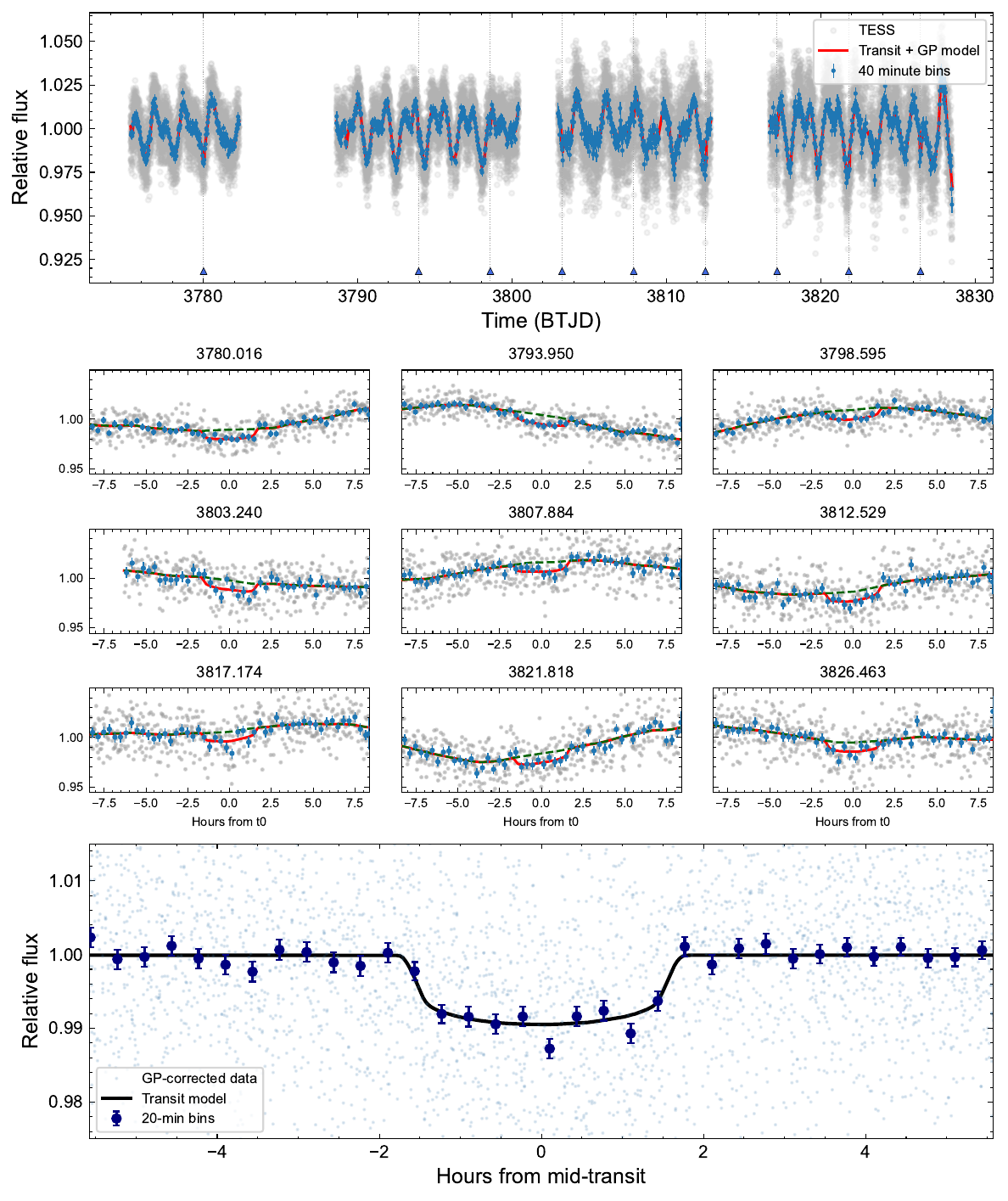}
    \caption{
        \textbf{TESS light curves of TIC~88297141.}
        \textit{Top:} TESS PDCSAP flux at 2-minute
        sampling, overlaid with 40-minute bins, from Sectors~91
        and~92. Starspot-induced variability, consistent with the
        star's 1.8-day rotation period, is the dominant signal.
        Markers indicate nine out of the 11 transits observed by TESS. 
        The remaining two, affected by data gaps, are omitted here for
        display but were included in the analyses. 
        \textit{Middle:} Individual TESS transits. The red line is the
        best-fit trend+transit model, while the dotted green line
        shows the trend without the transit.  Gray points are 2-minute
        PDCSAP flux, and the light blue points are binned
        to $20$-minute intervals.
        \textit{Bottom:} Phase-folded TESS transits, with the local
        spot-induced variability removed. Light blue points show the
        flattened flux measurements; a weighted binning at $20$-minute
        intervals yields the dark blue points, with the error bar
        showing the scatter in each bin. The black line is the
        best-fit transit model.
    }
    \label{fig:lightcurve}
\end{figure*}

We cleaned each light curve by removing quality-flagged data points,
masking NaNs, clipping flares via an interquartile range filter, and
normalizing by median flux.  We estimated stellar rotation periods
using a Lomb--Scargle periodogram \citep{Lomb1976, Scargle1982}, and
used the value to detrend the light curve with \texttt{wotan}
\citep{Hippke2019}, using a biweight filter with a $0.15\,P_{\rm rot}$
fiducial window length to balance overfitting the transit against
underfitting the stellar variability. Using
\texttt{TransitLeastSquares} \citep[TLS;][]{Hippke2019b}, we searched
the flattened light curves and retained candidates meeting a
top~$10\%$ Signal Detection Efficiency (SDE) threshold (SDE $\geq 10$)
across the sample. TIC~88297141 emerged in Sector~92 with $SDE = 10.52$, and was
independently recovered using a notch-filter search pipeline similar
to that of \citet{Rizzuto2017}. We masked the transits of TIC~88297141 to search 
for additional transiting companions, but found no significant periodic detection ($\mathrm{SDE} \ge 10$).
For the subsequent transit modeling, we used PDCSAP light curves from Sectors~91 and~92, which together
provide the most complete phase coverage of the transit signal. The
top panel of Figure~\ref{fig:lightcurve} shows the TESS data with
Sectors~91 and~92 stitched together. We measured the dominant spot-induced variability at a period of $\sim$\prot\,days. The
transiting candidate was measured at an orbital period of
$\sim$\period\,days. A simultaneous fit of the transit and variability
using Gaussian processes is described in Section~\ref{sec:juliet}.

The TESS SPOC pipeline detected this signal in its multi-sector search 
of Sectors 1--92, and it performed a suite of automated diagnostic tests 
as part of the data validation (DV) report for that detection.
TIC~88297141 passed all of these checks, including an odd/even
transit-depth comparison test, a weak secondary-eclipse test (no
statistically significant secondary event was detected; maximum
secondary MES $= 2.0\sigma$), and a ghost diagnostic test
\citep{Twicken2018}. A difference-image centroiding test placed the
transit source $2.78\pm2.70''$ from the target's catalog position, a
$1.03\sigma$ offset. This test's own $3\sigma$ confusion radius
($8.11''$) is larger than the $2\farcs6$ separation to the known bound
companion, so while it rules out more distant sources as the origin of
the signal, it cannot independently distinguish the primary from this
close companion. That distinction is instead established directly by
our higher-precision ground-based photometry
(Section~\ref{sec:groundphot}, Section~\ref{sec:fpscenarios}).

\subsection{Gaia Astrometry and Imaging}
\label{sec:gaia}

The \textit{Gaia} satellite \citep{GaiaMission2016} has measured
precise astrometry and photometry for over a billion stars, released
as \textit{Gaia} Data Release~3 (DR3; \citealt{GaiaDR32023,
Lindegren2021, Riello2021}).  TIC~88297141 was assigned the
\textit{Gaia} DR3 identifier 4109766661272060288, with a parallax of
$10.00\pm0.03$\,mas and a renormalized unit weight error (RUWE) of
$1.138$, consistent with a well-behaved, single-star astrometric
solution. Its brightness was measured in the $G$, $G_{\rm BP}$, and
$G_{\rm RP}$ bands ($G=14.29$, $G_{\rm BP}=15.91$, $G_{\rm RP}=13.04$;
\citealt{Riello2021}). Given its low Galactic latitude
($b=+5.1\degr$), the field surrounding TIC~88297141 is notably
crowded, motivating the ground-based and speckle imaging follow-up
described below.

\begin{figure}
\centering
\begin{minipage}{0.48\textwidth}
    \centering
    \includegraphics[width=\linewidth]{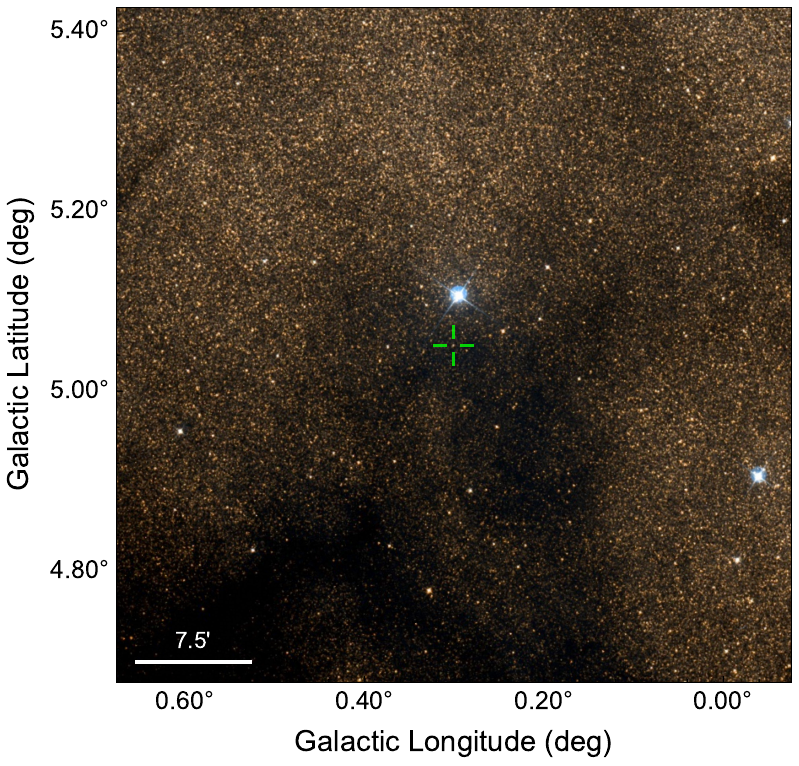}
\end{minipage}
\begin{minipage}{0.48\textwidth}
    \centering
    \includegraphics[width=\linewidth]{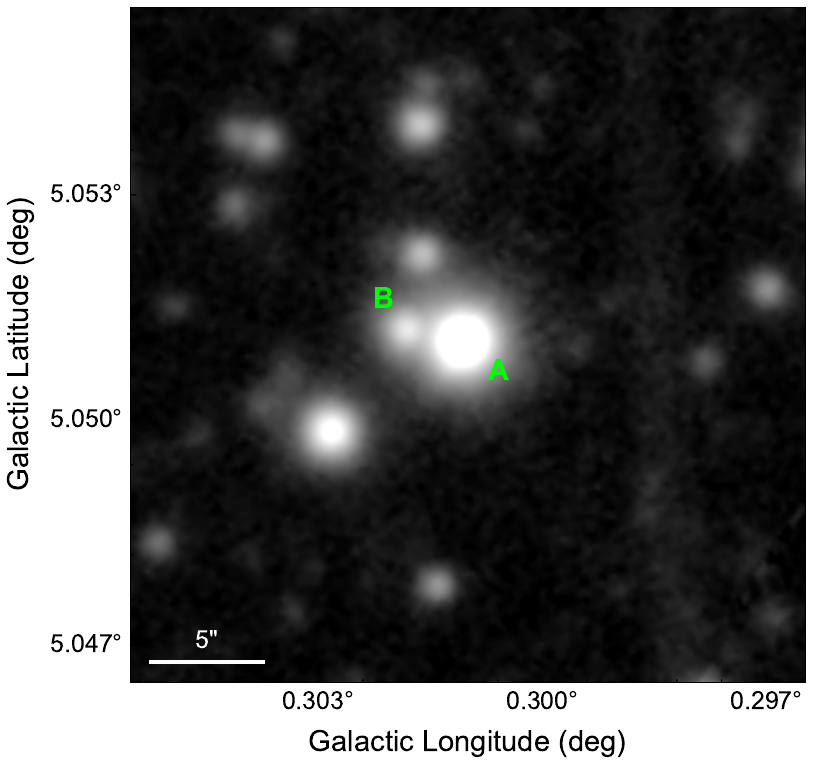}
\end{minipage}
\caption{\textbf{The scene.} \textit{Top:} $0.75\degr\times0.75\degr$
  field around TIC~88297141 from the Second Digitized Sky Survey
  (DSS2). \textit{Bottom:} Zoomed-in $30\arcsec\times30\arcsec$
  Pan-STARRS1 (PS1) DR1 $r$-band image \citep{chambers2016} showing
  the target (A) and the resolved companion (B), separated by
  $\sim2\farcs6$.}
\label{fig:field}
\end{figure}

The resolved source of immediate concern for our false-positive
analysis was Star~B $\equiv$ Gaia DR3\,4109766661238033408
(TIC\,1450801333; $G=16.42$), located $2\farcs568$ to the northeast of
the target (Figure~\ref{fig:field}). Star~B shares a parallax with 
TIC~88297141, and its proper motion is consistent with the two stars
being gravitationally bound, corresponding to a projected physical 
separation of only $\sim$260\,AU. Neither star shows evidence of an 
elevated RUWE, disfavoring an additional, unresolved astrometric 
companion around either component.

At the $\sim$1\arcmin\ angular resolution of the TESS photometric
aperture, if Star~B were itself an eclipsing binary, its diluted light
could be the source of the observed transit-like signal. No SPOC or
QLP light curve was produced for Star~B individually, precluding a
direct search for such a signal using our standard pipeline.
Ground-based, seeing-limited photometry (Section~\ref{sec:groundphot})
and high-resolution speckle imaging (Section~\ref{sec:speckle}) were
therefore necessary to test this possibility.

\subsection{Ground-Based Photometry}
\label{sec:groundphot}

We obtained ground-based photometric follow-up of TIC~88297141 from
several facilities. Most transit observations were obtained as part of
the TESS Follow-up Observing Program (TFOP) Sub-Group~1 (SG1)
initiative \citep{Collins2019}, using the network of 1\,m telescopes
equipped with Sinistro cameras operated by Las Cumbres Observatory
(LCOGT) \citep{Brown2013} at Siding Spring Observatory (SSO), Cerro
Tololo Inter-American Observatory (CTIO), and McDonald Observatory. In
addition, two transit observations were obtained independently of the
TFOP program using the CTIO 1\,m (Section~\ref{sec:CTIO}), and we
obtained further observations using the Swope 1\,m telescope at Las
Campanas Observatory in Chile
(Section~\ref{sec:swope}). Images taken via the LCOGT network were
calibrated using the standard LCOGT \texttt{BANZAI} pipeline
\citep{McCully2018}. All ground-based data were reduced using
\texttt{AstroImageJ} \citep{Collins2017}.  We selected comparison
stars and apertures to minimize the dispersion
(Figure~\ref{fig:groundmosaic}).  Analysis of our ground-based
photometry confirmed that the transit event is on-target, with the
signal in one of our observations localized within a $1.2''$ aperture.
Table~\ref{tab:groundphot} summarizes the aperture sizes for all the
ground-based data. 

\subsubsection{Swope 1\,m (Las Campanas Observatory)}
\label{sec:swope}

We observed a predicted egress of TIC~88297141 on UT~2026~March~24
using an e2v 4k $\times$ 4k CCD camera on the Swope 1\,m telescope through the Sloan $r'$ filter. A total of 201 images were
obtained over $\sim$295\,minutes, with exposure times varying between
20 and 40\,s, yielding a mean cadence of $\sim$89\,s. 

\subsubsection{Siding Spring Observatory 1\,m}
\label{sec:sso}

We observed a full predicted transit of TIC~88297141 on
UT~2026~April~25 using the Sinistro camera on the LCOGT 1\,m telescope
at Siding Spring Observatory, as part of the TFOP SG1 program, through
the SDSS $i'$ filter, with a cadence of $\sim$90\,s. 

\subsubsection{CTIO}
\label{sec:CTIO}

We observed a predicted transit ingress of TIC~88297141 on
UT~2026~April~16 using the Sinistro camera on the LCOGT 1\,m telescope
at CTIO, as part of the TFOP SG1 program, simultaneously in SDSS $g'$
and SDSS $i'$ filters, with a cadence of $\sim$49\,s.

In addition, a full predicted transit was observed independently of
the TFOP program on UT~2026~June~24 using the Sinistro camera on the
CTIO 1\,m telescope, simultaneously in SDSS $g'$ and Pan-STARRS $z_s$
filters, with a cadence of $\sim$70\,s in both bands. Photometry was
extracted using the \texttt{prose} package \citep{prose}, with
comparison stars selected automatically following the
differential-photometry algorithm of \citet{Broeg2005}.

\subsubsection{McDonald} 
\label{sec:McD}

We observed a predicted transit ingress of TIC~88297141 on 
UT~2026~April~16 using Sinistro cameras on two LCOGT 1\,m telescopes 
at McDonald Observatory, as part of the TFOP SG1 program, 
simultaneously in SDSS $i'$ and $g'$ filters, with cadences of 
$\sim$90\,s and $\sim$127\,s, respectively.

\input{apertures.tex}

\begin{figure*}[t]
    \centering
    \includegraphics[width=\textwidth]{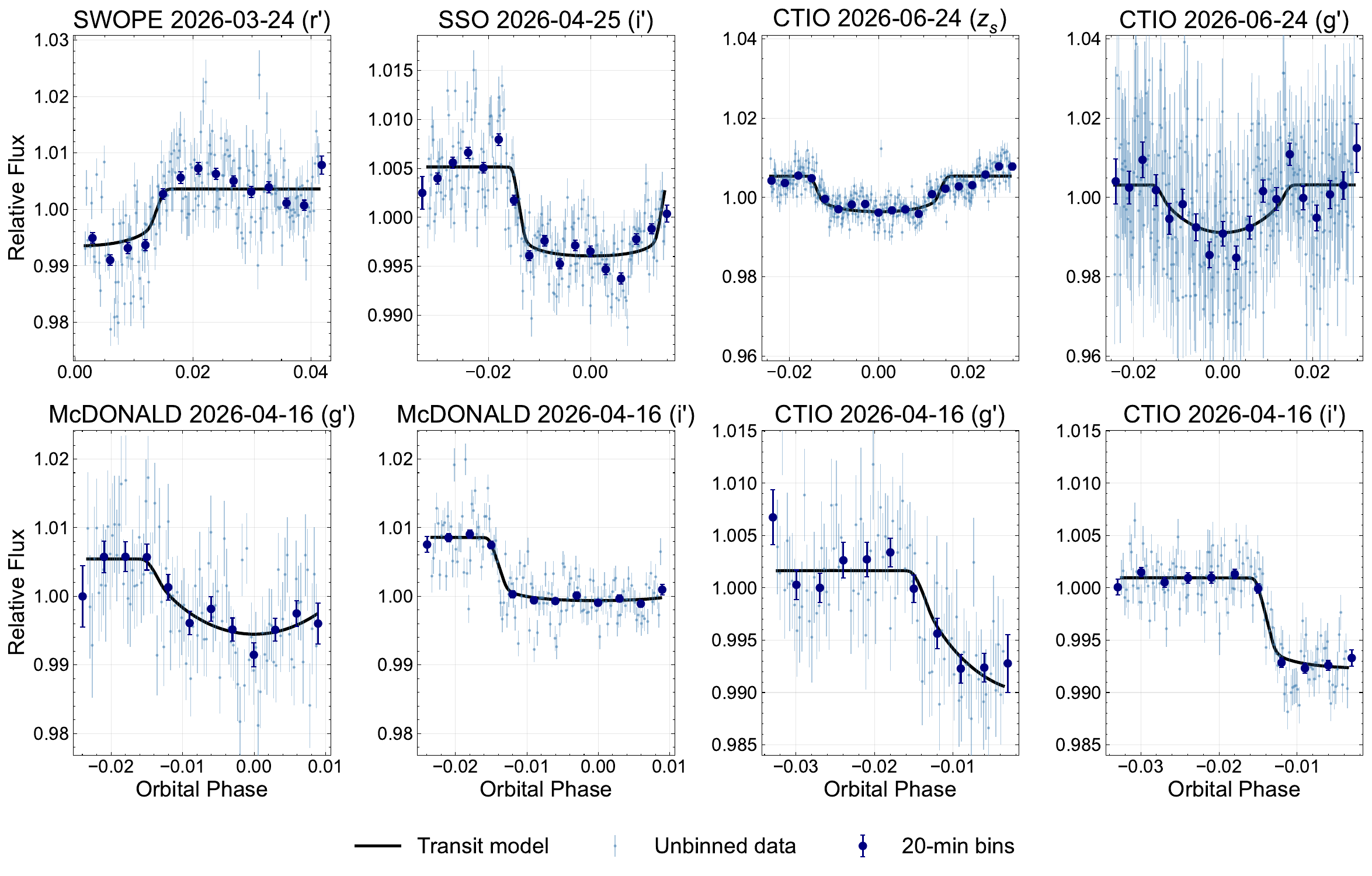}
    \caption{\textbf{Seeing-limited photometry.} Detrended
    ground-based transit light curves with the best-fit transit model (black)
    overplotted on the unbinned (light blue) and 20-minute binned
    (navy) data. For each observing sequence, the $y$-axis range 
    is intentionally matched across its observed filters (CTIO $g'$ and $z_s$ on 2026-06-24; McDonald $g'$ and $i'$, and CTIO $g'$ and $z_s$, on 2026-04-16), 
    allowing the relative scatter between bands to be compared without introducing differences from independently scaled axes. The three $g$-band
    light curves (CTIO 2026-06-24, CTIO 2026-04-16, McDonald
    2026-04-16) show larger scatter than their same-night redder-band
    counterparts.}
    \label{fig:groundmosaic}
\end{figure*}

Because TIC~88297141 is a cool M dwarf ($T_{\rm eff}\approx\teff$~K),
its flux decreases toward bluer wavelengths, resulting in lower
per-point signal-to-noise in our $g'$-band light curves than in the
redder bandpasses observed on the same night
(Figure~\ref{fig:groundmosaic}). In addition, stellar limb darkening
is stronger at shorter wavelengths, producing intrinsically more
rounded transit light curves \citep{winn2010}. In our observations,
however, the increased point-to-point scatter in the $g'$-band data
appears to be the dominant effect, making the ingress and egress less
well defined than in the corresponding same-night $i'$ or $z_{s}$-band
observations. We also found the CTIO $g'$-band light curve collected
in June to be sensitive to the choice of comparison stars, which is
consistent with the observed differences arising primarily from
photometric noise. We discuss the implications of the multi-band
photometry for false-positive exclusion in Section~\ref{sec:mcph}.

\subsubsection{Contamination}
\label{sec:contamination}

The primary source of contamination in our ground-based photometry is
the resolved binary companion (Section~\ref{sec:gaia}).  For each
observation, we calculated the photometric dilution factor using the
reported aperture radius and image PSF FWHM.  We modeled both stars as
circular Gaussian PSFs, setting the companion's relative flux,
separation, and position angle to our measured values.  Integrating
both PSFs over the photometric aperture yielded the dilution factor
\begin{equation}
D =
\frac{F_{\rm target}}
     {F_{\rm target}+F_{\rm neighbor}}.
\end{equation}
The resulting values, listed in Table~\ref{tab:groundphot},
were held fixed in the transit fits
(Section~\ref{sec:juliet}). TESS PDCSAP photometry is already corrected
for crowding by SPOC, so no additional dilution was applied to the
TESS data. Because the PSF is approximated as Gaussian, they were regarded 
as approximations.

\subsection{Speckle Imaging}
\label{sec:speckle}

\begin{figure}[htbp]
    \centering
    \includegraphics[width=\columnwidth]{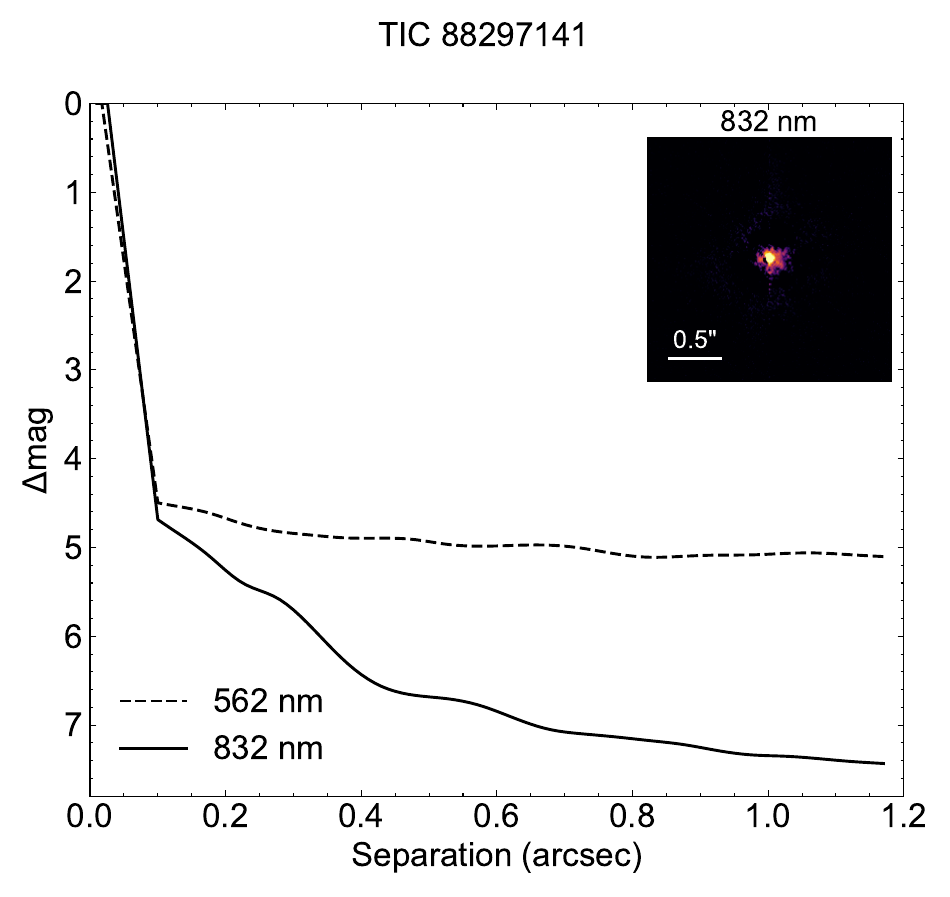}
    \caption{\textbf{Speckle imaging of TIC~88297141.} $5\sigma$ magnitude 
    contrast curves from Gemini-South Zorro imaging Contrast were derived from 
    the triple correlation reconstructed images \citep{Howell2011}, with the 562~nm (blue) and 832~nm
    (red) channels overplotted. The 832~nm curve is more constraining
    than the 562~nm curve at larger separations, with no evidence for
    companions within $1\farcs2$.}
    \label{fig:zorro_contrast}
\end{figure}

To search for unresolved stellar companions within a few tenths of an
arcsecond of TIC~88297141 that could be the source of the transit
signal, we obtained speckle imaging on UT~2026~April~30 using the
Zorro instrument on the Gemini-South 8\,m telescope \citep{Scott2021}.
Zorro employs a dichroic beamsplitter to simultaneously image in blue
and red channels, here using $562$\,nm and $832$\,nm filters,
respectively, with a plate scale of $0.01\arcsec$\,pixel$^{-1}$. The images 
obtained were subjected to Fourier analysis in our standard reduction pipeline 
(see \citet{Howell2011}). We detected no additional close-in
stellar companions (Figure~\ref{fig:zorro_contrast}), with contrast
limits of $\Delta m = 4.93$\,mag and $\Delta m = 6.68$\,mag at a
separation of $0.5\arcsec$ in the $562$\,nm and $832$\,nm bands,
respectively. The known resolved binary companion discussed in
Section~\ref{sec:gaia} lies at a separation well beyond the field of
view probed by this speckle observation, and is therefore not
constrained by these contrast curves.

\subsection{Radial Velocities}
\label{sec:rv}

We obtained radial velocity measurements of TIC~88297141 using the
Planet Finder Spectrograph (PFS) \citep{Crane2006, Crane2008, Crane2010}, a high-resolution echelle
spectrograph on the Magellan~II (Clay) 6.5\,m telescope at Las
Campanas Observatory. PFS was used with the 0.3\arcsec\ slit, which delivers a
default resolving power of $R \approx 130{,}000$ under standard
$1\times2$ pixel binning. Due to the faint magnitude of the target, we
instead used $3\times3$ binning, which reduces the effective resolving
power to $R \approx 110{,}000$. Observations were
taken using the iodine-cell technique for precise wavelength
calibration \citep{Butler1996}. During this observing run, the star
was observed on three nights (UT~2026~April~26, 27, and 29), together
with a separate 6-exposure, iodine-free template observation
(UT~2026~April~28); a fourth epoch was obtained during a subsequent
run on UT~2026~June~8, together spanning a total baseline of
$\sim$43\,days. With $\sim$65 photons pixel$^{-1}$ in the iodine
region and a per-pixel signal-to-noise ratio of $\sim$8 in individual
exposures, TIC~88297141 is among the faintest stars observed with PFS
to date. Before any correction for the intra-epoch scatter,
the formal uncertainties on the RVs were $\sim$16--20\,m\,s$^{-1}$,
reflecting this reduced photon flux. At each epoch, three consecutive exposures were combined into a single
inverse-variance-weighted mean radial velocity, with the propagated
uncertainty scaled by $\sqrt{\chi_\nu^2}$ whenever the reduced
chi-squared of the three constituent exposures exceeded unity, thereby
accounting for excess intra-epoch scatter. Given that the three 
exposures per epoch span less than an hour, a small fraction of the
$\sim1.8$ day rotation period, the intra-epoch scatter
is unlikely to arise from rotational activity. We thus attribute it to
reduced reliability of per-exposure velocity extraction in this
photon-starved (S/N$\sim8$) regime. The resulting radial velocities
and their uncertainties are listed in Table~\ref{tab:rv}.

The four PFS epochs show a large epoch-to-epoch scatter, with a sample RMS of
$\approx 320~\mathrm{m\,s^{-1}}$, much larger than the per-epoch uncertainties. We tested whether 
this scatter could be explained by stellar activity via a rough estimation of the spot-induced RV signal
using the simple spot model of \citet{Aigrain2012}, which relates the RV perturbation to
the photometric spot modulation and the equatorial rotation velocity. The TESS light curves show
a peak-to-peak spot modulation of roughly $4$--$5\%$, and with
$v_{\rm eq}=20.4~\mathrm{km\,s^{-1}}$ (Section~\ref{sec:Prot}), we expect activity-induced RV
variations of a few hundred $\mathrm{m\,s^{-1}}$ in an idealized single-spot geometry, with an
RMS of roughly $230$--$320~\mathrm{m\,s^{-1}}$ depending on the TESS sector. The observed
scatter is therefore comparable to what the rapidly rotating host is expected to
produce.

Our estimates assume an idealized spot geometry, use TESS
photometry taken about a year before the PFS epochs, and rest on only four epochs. We
thus conclude only that the observed scatter is consistent with stellar activity. Since activity of this 
amplitude can mimic or mask a planetary signal in sparsely sampled RV
data, we also tested whether a planetary signal is statistically supported in Section~\ref{sec:pfsfp}.

\input{PFS_rv_table.tex}

\subsection{SOAR Broadband Imaging}
\label{sec:soar}

Because TIC~88297141 lies in a crowded field, with a resolved comoving
stellar companion $\sim2\farcs6$ away (Section~\ref{sec:gaia}), we
obtained resolved $g$, $r$, and $i$-band imaging of the system using
the Goodman High Throughput Spectrograph on the SOAR 4.1\,m telescope
to help constrain its broadband spectral energy distribution (SED).
The target is blended in existing wide-field catalogs: Pan-STARRS1
(PS1) reports three sources within $\ang{;;5}$, while SOAR resolves
the field into one dominant source (the target) and two fainter,
well-separated neighbors. We fit the three sources simultaneously
with Moffat profiles at their forced positions, using the fit solely
to subtract the two neighbors before measuring the target's flux on
the neighbor-subtracted frame. The resulting photometry was
calibrated directly against other PS1 stars in the same field, each
with no PS1 neighbor within $\ang{;;3}$, via a zero point and linear
color term. The deblended photometry of the primary
(Table~\ref{tab:stellar_params}) is consistent with archival
Pan-STARRS1 photometry within measurement uncertainties.

\section{The Star}
\label{sec:star}

\input{stellar_properties.tex}

\subsection{Stellar Parameters}
\label{sec:sed}

Table~\ref{tab:stellar_params} lists literature and derived properties
for TIC~88297141A and its bound companion. We determined the primary's
stellar parameters using \texttt{AstroARIADNE} \citep{Vines2022},
adopting the BT-Settl atmosphere grid \citep{Allard2012}.
\texttt{AstroARIADNE} fits the observed spectral energy distribution
(SED) against a grid of stellar atmosphere models for six parameters,
$T_{\rm eff}$, $R_*$, $A_V$, distance, $\log\,g$, and
$\mathrm{[Fe/H]}$, with a distance prior drawn from
\citet{BailerJones2021}. We adopted uniform priors on $T_{\rm eff}$
($2000$--$8000$~K), $\mathrm{[Fe/H]}$ ($-0.3$ to $0.3$~dex), $R_*$
($0.1$--$1.5\,R_\odot$), and $A_V$ ($0.0$--$0.2$~mag), and a normal
prior on $\log\,g$ ($\mathcal{N}(4.5, 0.5)$).  
We fit ten broadband magnitudes: Pan-STARRS1 $g,r,i,z,y$, Gaia DR2
$BP$/$G$/$RP$, SkyMapper $r$, and SDSS $i$; we excluded 2MASS and WISE
photometry, since TIC~88297141 is flagged as crowded in both catalogs,
consistent with its resolved $\sim2\farcs6$ companion. Of the six fitted
parameters, only $T_{\rm eff}$ and $R_*$ are well constrained ($T_{\rm
eff}=3168^{+36}_{-71}$~K, $R_*=0.7375^{+0.0243}_{-0.0234}\,R_\odot$);
$\mathrm{[Fe/H]}$ is weakly constrained but the median value is still
reported in Table~\ref{tab:stellar_params}. We derived $\log\,g$ by
performing Monte Carlo sampling, drawing random samples combining the
SED-derived radius with the evolutionary mass derived from Feiden
tracks (Section~\ref{sec:mass}), propagating their uncertainties.
Additionally, the fit reports an unconstrained stellar age of
$11.5^{+4569}_{-4.5}$~Myr. We thus explore the age of the star in
Section~\ref{sec:age} and Section~\ref{sec:membership}.

\subsection{Lithium}
\label{sec:age}

Fully convective mid-M dwarfs mix photospheric material down to the
core efficiently, so once the core reaches the lithium-burning
temperature ($\sim 2.5\times10^6$~K), the star's entire lithium supply
is depleted on a timescale of a few Myr \citep{bildsten1997}.  This
mass-dependent depletion timescale is the physical basis of the
lithium depletion boundary technique used to age-date young clusters
\citep{basri1996}, and it means that any lithium still detectable in a
star of TIC~88297141's mass is itself informative about the system's
youth.

The co-added iodine-free PFS spectra of TIC~88297141 show the 6708~\AA\ lithium
doublet in absorption. We measured the line's equivalent width from a
continuum-normalized Gaussian fit, propagating uncertainty via Monte
Carlo resampling of the continuum noise, and obtained EW~$=
345.3^{+36.1}_{-26.6}$~m\AA. Feeding this EW, together with the
SED-derived effective temperature ($T_{\rm eff} = 3168$~K), into
\texttt{eagles} \citep{jeffries2023, weaver} yields a most-probable
age of $18.2^{+3.2}_{-15.7}$~Myr. The posterior's lower bound is
essentially unconstrained, but the upper bound places a firm
constraint on the stellar age, confirming its youth.

\subsection{Rotation}
\label{sec:Prot}

We measured the rotation period from the TESS PDCSAP light curve using
the Lomb--Scargle periodogram \citep{Lomb1976, Scargle1982}. We derived 
the uncertainty on the best period using Equation~1 in \citet{Boyle2025rot}. We adopted
two Fourier terms in the periodogram model because the rotational modulation is non-sinusoidal 
(double-humped). We measured $P_{\rm rot} =
1.832 \pm 0.022$\,d. This period is also consistent 
with the independent value reported by the TESS All-Sky Rotation Survey \citep[TARS;][]{Boyle2026}, who derive
$P_{\rm rot} = 1.8192$\,d for this target. Combining the
rotation period with the SED-derived radius from
Section~\ref{sec:sed}, we derived an equatorial velocity of $v_{\rm eq} = 20.42^{+0.68}_{-0.73}$\,km\,s$^{-1}$. We use this derived $v_{\rm eq}$ in
Section~\ref{sec:discussion} to predict an approximate expected
Rossiter--McLaughlin (RM) amplitude. 

As stars get older, their rotation rates slow due to magnetic braking
\citep{Skumanich1972, WeberDavis1967}, a relation well calibrated for
Sun-like stars. For fully convective M dwarfs, however, the timescale
for this spin-down is poorly measured: evidence from the MEarth survey
showed such stars can remain in a rapidly rotating ($P_{\rm rot}
\lesssim 10$\,d) state for an extended period before an abrupt
transition to slow rotation \citep{Newton_2016}. Recent work has
revisited this transition without a definitive resolution
\citep{Newton_2018, LuCurtis2022, PassCharbonneau2024}.  The closest
empirical anchor comes from the 4\,Gyr cluster M67, where
\citet{Dungee2022} found a converged, single-valued rotation sequence
extending through mid-M dwarfs. This sequence, however, does not
directly probe the lower-mass regime that includes TIC~88297141\,A.
The rotation period of TIC~88297141\,A thus does not yield a precise
age, but it is inconsistent with the star being an old, slowly
rotating field M dwarf, and is consistent with membership in the young
Sco-Cen population established independently via lithium
(Section~\ref{sec:age}) and kinematics (Section~\ref{sec:membership}).

\subsection{The Companion}
\label{sec:companion}

Star~B (Section~\ref{sec:gaia}) is more likely
a physical companion than a chance alignment:
its parallax is consistent with that of the primary,
its proper motion is consistent with the two stars 
being gravitationally bound, and its photometry falls 
on the same isochrone. At an angular separation of $\sim 2\farcs6$,
corresponding to a projected separation of $\sim260$\,AU, these
properties support an interpretation of Star~B as a bound stellar
companion.

The companion is fainter than the target ($G=16.42$), and we constrain
its stellar parameters through its broadband photometry as portrayed
in Table~\ref{tab:stellar_params}. We fit the companion's SED using
the same \texttt{AstroARIADNE} setup and priors as for the primary
(Section~\ref{sec:sed}). Because the pair is separated by only
$2\farcs6$, most surveys (2MASS, AllWISE, SkyMapper, APASS) return a
single blended source dominated by TIC~88297141A; only Pan-STARRS1
resolves the companion. We therefore fit six bands: Pan-STARRS1
$i,z,y$ and Gaia DR2 $BP$/$G$/$RP$, inflating the $BP$/$RP$
uncertainties to account for residual flux contamination from the
primary in the \textit{Gaia} window. As with the primary, only $T_{\rm
eff}$ and $R_*$ are well constrained ($T_{\rm
eff}=2843^{+142}_{-183}$~K, $R_*=0.394^{+0.10}_{-0.08}\,R_\odot$); we
treated and reported $\log\,g$ and $\mathrm{[Fe/H]}$ as in
Section~\ref{sec:sed}.

\subsection{Sco-Cen Membership}
\label{sec:membership}

We assessed TIC~88297141's Sco-Cen membership by studying previous
literature clustering analyses and by searching the system's local
neighborhood for co-moving, co-spatial, and co-eval stars.

Understanding of Sco-Cen's structure has advanced since Hipparcos
\citep{dezeeuw1999, Preibisch2008}.  In particular, Gaia positions and
velocities have enabled the discovery of dozens of subgroups beyond
the traditional Upper Scorpius, Upper Centaurus Lupus, and Lower
Centaurus Crux boundaries \citep{Kerr2021,Ratzenbock2023a}.  These
subgroups enable reconstructing the association's detailed star
formation history in space and time \citep[e.g.,][]{Ratzenbock2023b}.

TIC~88297141 was reported as a Sco-Cen member by \citet{Kerr2021} and
\citet{Ratzenbock2023a}.  \citet{Hunt2023} additionally reported
TIC~88297141 as a member of HSC\,2976 ($N$=21, $\log
t$=7.40$^{+0.49}_{-0.67}$, $A_V$=0.84$^{+0.42}_{-0.62}$);
\citet{Kounkel2019} placed the system in a region associated with
Sco-Cen.  Since the latter two studies did not include detailed
treatment of Sco-Cen's substructure, we omit them from further
discussion.

\begin{figure*}[!t]
    \centering
    \includegraphics[width=0.98\textwidth]{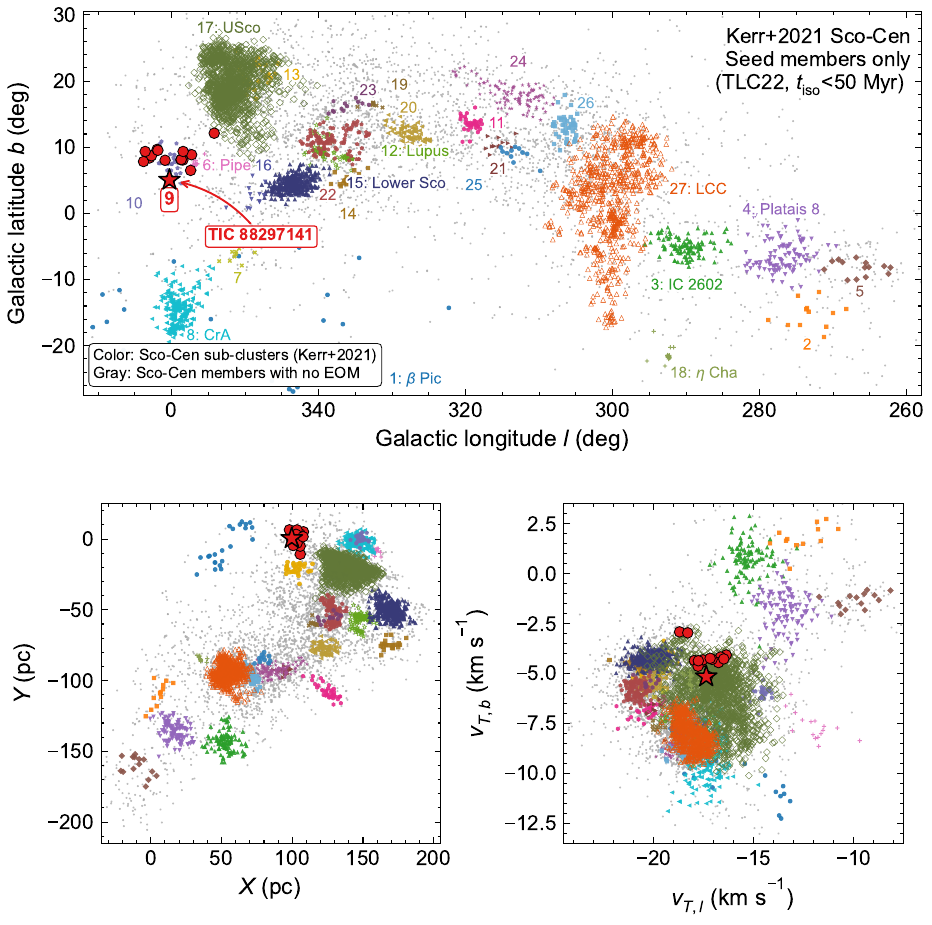}
    \vspace{-0.4cm}
      \caption{{\bf TIC~88297141 resides in a low-density Sco-Cen
      subcluster.}  This Figure reproduces \citet{Kerr2021} Figure~18;
      the points show ``seed'' members from \citet{Kerr2021} with
      at least a 10\% probability of having an isochrone age $t_{\rm
      iso} < 50$\,Myr.  Points in color are associated with an excess of mass
      (EOM), implying the existence of a subcluster.  Points in gray
      (51.6\% of the 7{,}138 TLC~22 members) are Sco-Cen members that fell
      below the \citet{Kerr2021} HDBSCAN fragmentation threshold and which
      are not associated with any subcluster.}
      \label{fig:kerr}
\end{figure*}

Figure~\ref{fig:kerr} shows the position and velocity of TIC~88297141
relative to Sco-Cen.  TIC~88297141 resides southeast of Upper Sco, in
a comparatively low-density region of position and velocity space.  In
Figure~\ref{fig:kerr} we compare the system against the
\citet{Kerr2021} clustering results.  In detail, \citet{Kerr2021}
reported TIC~88297141 as one of the seed members of an $N$=15 star
subgroup labeled excess of mass (``EOM'') 9, and derived an ensemble
isochrone age $t_{\rm iso,ens}$=15.6$\pm$1.6\,Myr for the group based
on the PARSEC v1.2S isochrones \citep{Bressan2012,Chen2015}, with the
\citet{Lallement2019} dereddening maps.  \citet{Kerr2021} reported an
isochrone age for the planet-hosting primary star of $t_{\rm
iso,A}$=15.3$^{+3.2}_{-5.7}$\,Myr.   These ``seed'' members are stars
with isochrone ages $t_{\rm iso}$$<$50\,Myr, excluding most FGK
members \citep{Kerr2021}.   In comparison, \citet{Ratzenbock2023a}
reported TIC~88297141 as a member of $\eta$~Lup (SigMA~14), an $N$=769
star, 14.8$\pm$0.7\,Myr subgroup primarily located in UCL, near Kerr's
EOM22.  However, there are two key issues with the
\citet{Ratzenbock2023a} subgroup assignment. First, the reported
stability (34\%) is low.  More important, the extension of $\eta$~Lup
to this region of position and velocity space (see Figure~E.2 of
\citealt{Ratzenbock2023a}) seems likely to be an overreach of the
SigMA clustering algorithm: in $Y$~vs.~$X$ in particular, the region
near TIC~88297141 is grouped with the core of $\eta$~Lup nearly 50\,pc
distant.  Fundamentally, the differences between \citet{Kerr2021} and
\citet{Ratzenbock2023a} are produced by differences in clustering
methodology: the \citet{Ratzenbock2023a} SigMA method yielded a flat
grouping that forced every source in the input spatial box to be part
of a subgroup; the \citet{Kerr2021} application of HDBSCAN produced a
nested three-level hierarchical clustering
(TLC$\rightarrow$EOM$\rightarrow$leaf), with a fragmentation floor of
25\,pc in the spatial axes and 4.2\,km\,s$^{-1}$ in the velocity axes,
and a minimum cluster size of 10 photometrically young stars.  The
\citet{Kerr2021} delineation of TLC22 EOM9 as a standalone subgroup
seems to better mirror the spatial structure of the pre-main-sequence
stars in this region; we therefore adopt it as our default group
membership assignment for TIC~88297141.

As an independent test on the group assignment, we used
\texttt{Comove} \citep{Tofflemire2021} to query Gaia DR3
\citep{GaiaDR32023} for comoving stars in the local spatial
neighborhood of TIC~88297141.  Within 20 parsecs, 38 stars co-move
with TIC~88297141\,A at tangential velocity $v_{\rm
T}$$<$1\,km\,s$^{-1}$.  Comparing against empirical cluster isochrones
shows that most of these comovers are photometrically consistent with
the isochrone ages listed above.  \citet{Ratzenbock2023a} labeled 27
of these 38 stars as being part of $\eta$~Lup (though as noted above,
we contest the extension of $\eta$~Lup to this spatiokinematic
region).  \citet{Ratzenbock2023a} assigned two stars in this volume to
other groups, and excluded nine as field stars.  Combining their
Tables~1 and 2, \citet{Kerr2021} labeled 21 (13+8) of the 38
co-spatial, co-moving stars as being part of TLC~22 EOM~9; nine (7+2)
as being in EOM~$-1$ (unassigned Sco-Cen), and the excluded eight as
field stars.

\subsection{Mass Constraints} 
\label{sec:mass}

\texttt{AstroARIADNE}'s isochrone-based mass assumes standard,
non-magnetic MIST evolutionary models. However, \citet{Feiden2016} has
shown that considering magnetic inhibition of convection is
increasingly important to reproduce the radii of fully convective
young M dwarfs. This result motivates a comparison against a mass estimate that does not
rely on stellar evolutionary models. First, we use the posterior samples
of the stellar density from the transit fit in Section~\ref{sec:transit},
which are independent of stellar models, and
combine them with the SED-fit radius recovered in Section~\ref{sec:sed}
to obtain an empirically derived mass (Table~\ref{tab:stellar_params}). We propagate both posteriors,
finding $M_{\star} = 0.307^{+0.055}_{-0.072}\,M_\odot$, in good
agreement with \texttt{AstroARIADNE}'s own isochrone mass
($0.3005\,M_\odot$). Because the transit fit assumes a circular orbit, this mass is
conditioned on $e=0$: a nonzero eccentricity would bias the
transit-derived stellar density, and hence this mass
\citep[the photoeccentric effect;][]{Kipping2010, Dawson2012}.

We derived theoretical stellar masses by comparing our SED-derived
$T_{\rm eff}$ and $R_\star$ against the \citet{Feiden2016} magnetic
grids. For each Monte Carlo draw of the observed parameters, we
additionally drew a system age from $\mathcal{N}(15.6, 1.6)$~Myr,
propagating the age uncertainty alongside $T_{\rm eff}$ and $R_\star$,
and performed a $\chi^2$ minimization along the corresponding
isochrone to obtain a theoretical mass estimate for that draw. This
Monte Carlo isochrone fitting therefore propagates both observational
and age uncertainty into the resulting mass predictions, yielding
$M_{\star,\mathrm{iso}} = 0.352 \pm 0.028\,M_\odot$, consistent with
our empirically derived mass within $1\sigma$. 

We combined the empirically derived mass with
the SED-derived radius to solve for the stellar age on the Feiden
grid.  We obtained an age of $11.51^{+2.02}_{-3.37}$\,Myr, consistent
with both the lithium age derived in Section~\ref{sec:age} and the
\citet{Kerr2021} target isochrone age in Section~\ref{sec:membership}
within uncertainties. Since the derived age also inherits the circular-orbit assumption
described above, we treat it as a consistency check alongside the lithium and
cluster isochrone ages.

\section{Transit Analysis}
\label{sec:transit}

\subsection{Joint Light Curve Transit Fit}
\label{sec:juliet}

We performed a joint fit of the transit and stellar variability signal
using \texttt{juliet} \citep{Espinoza2019}, which combines the
\texttt{batman} \citep{batman} transit model with a Gaussian Process
(GP) framework built on \texttt{celerite} \citep{ForemanMackey2017}
and explores the parameter space using the \texttt{dynesty} nested
sampler \citep{Speagle2020}.  We fit the two TESS sectors
simultaneously with eight ground-based light curves: the partial
(egress-only) Swope 1\,m transit, the full LCOGT/SSO 1\,m transit, two
independent epochs of LCOGT/CTIO 1\,m photometry ($g$- and $z$-band
observations in June, and separate $g$- and $i$-band observations in
April), and LCOGT/McDonald 1\,m $g$- and $i$-band photometry. 

The transit component was parameterized by the orbital period ($P$),
time of transit center ($t_0$), planet-to-star radius ratio
($R_p/R_*$), impact parameter ($b$), and stellar density ($\rho_*$),
from which \texttt{juliet} internally derives $a/R_*$ via Kepler's
third law. The orbit was fixed to be circular ($e=0$). Because the
transit and orbital parameters describe the planet itself, they were
treated as global parameters shared across all nine datasets in the
joint fit. We placed Gaussian priors on $P$ and $t_0$, centered on the
values recovered from our transit-detection pipeline
(Section~\ref{sec:tess}; $P = 4.64423$\,d, $t_0 = 3803.24126$\,BTJD)
with widths of $\sigma = 0.01$\,d and $0.1$\,d, respectively --
conservatively wide compared to the formal detection-pipeline
uncertainties, so as not to over-constrain the joint fit. $R_p/R_*$
and $b$ were given wide, uninformative uniform priors
($\mathcal{U}(0,1)$ and $\mathcal{U}(0,1.1)$, respectively), so that
the transit shape is constrained by the combined photometry rather
than by an assumed prior from any single dataset.

Limb darkening was described using the $(q_1, q_2)$ parameterization
of \citet{Kipping2013}, sampled independently for TESS, and shared
within filter-matched groups among the ground-based data: all $g$-band
epochs (CTIO-June, CTIO-April, McDonald) shared a single $(q_1,q_2)$
pair, as did the $i$/$z$-adjacent epochs (SSO, McDonald-$i$,
CTIO-April-$i$), while Swope and CTIO-$z$, which had no filter partner
among the remaining datasets, were each given independent
limb-darkening parameters.

The photometric dilution factor for each ground-based light curve was
fixed during the fit, computed from the properties of the known
resolved companion (Section~\ref{sec:sed}) following the
aperture-overlap procedure described in Section~\ref{sec:groundphot}. 

A per-instrument flux offset and white-noise jitter term were treated
as local, instrument-specific nuisance parameters throughout. For the
eight ground-based datasets, which individually span too short a
baseline to constrain stellar rotational variability, we replaced GP
detrending with a second-order polynomial in time (mean-subtracted
linear and quadratic terms), fit independently per instrument; any
residual correlated noise in these datasets is absorbed jointly by
this polynomial term and by the per-instrument jitter. Stellar
rotational variability in the TESS photometry itself was instead
modeled with a quasi-periodic (QP) GP kernel, following the
\texttt{celerite}-compatible formulation of \citet{ForemanMackey2017}
as implemented in \texttt{juliet}:

\begin{equation}
    k(\tau) = \frac{B}{2+C}\, e^{-\tau/L}
    \left[\cos\left(\frac{2\pi\tau}{P_{\rm rot}}\right) + (1+C)\right],
    \label{eq:qpkernel}
\end{equation}

where $\tau = |t_{j} - t_{i}|$ is the time lag between two photometric
observations, $B$ and $C$ set the covariance amplitude and the
relative weight of the oscillatory versus non-oscillatory terms, $L$
is the exponential decay timescale, and $P_{\rm rot}$ is the
oscillation period, corresponding to the stellar rotation period. We
placed a Gaussian prior on $P_{\rm rot}$ centered on 1.83\,d,
recovered following the methodology explained in
Section~\ref{sec:Prot}.  The remaining transit and GP parameters were
given uniform or log-uniform priors over physically motivated ranges
(Table~\ref{tab:tab5}).

We ran the nested sampler with 750 live points using multi-ellipsoidal
decomposition (\texttt{bound = `multi'}) and random-slice
(\texttt{rslice}) sampling, adopting a nominal stopping criterion of
$\Delta\ln Z = 0.5$. In practice the sampler
continued well past this threshold, with the final estimated remaining
evidence reaching $\Delta\ln Z \approx 0.001$, indicating that
convergence was not compromised by the looser nominal setting. 

Figures~\ref{fig:lightcurve} and \ref{fig:groundmosaic}, previously
introduced in Section~\ref{sec:tess} and Section~\ref{sec:groundphot}
respectively, show this same photometry together with the resulting
best-fit transit model from the joint fit described above: the transit
shape and depth are reproduced consistently across all nine datasets.

\begin{figure}[t]
    \centering
    \includegraphics[width=\columnwidth]{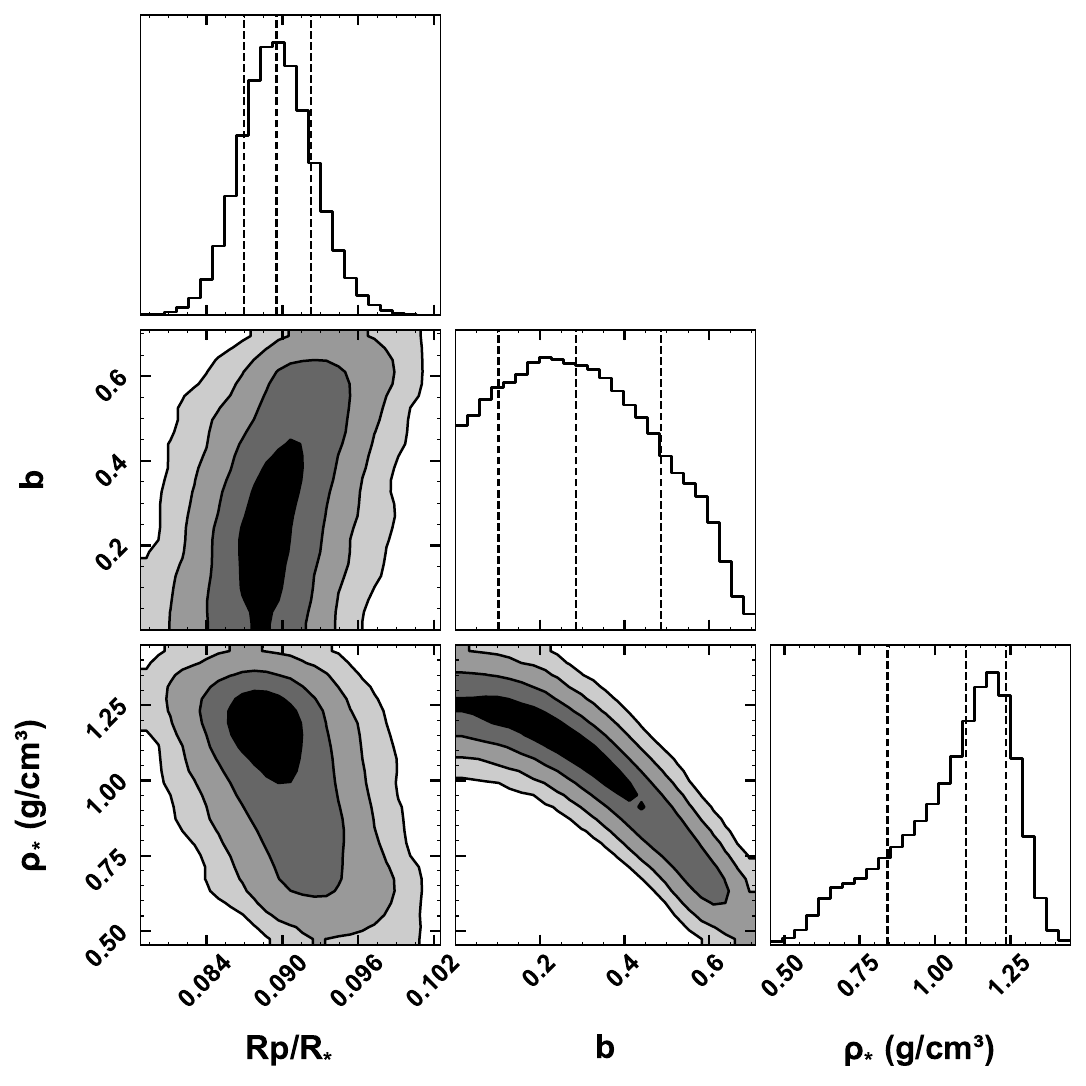}
    \caption{\textbf{Selected system parameters.} Posterior probabilities
    of the planet-to-star size ratio, impact parameter, and stellar
    density from the joint \texttt{juliet} fit of the TESS and
    ground-based photometry.  Dashed lines on the 1D histograms mark
    the 16th, 50th (median), and 84th percentiles of each posterior
    distribution. Contours are shown at $1\sigma$, $2\sigma$,
    $3\sigma$, and $4\sigma$ confidence. This plot was made using
    \texttt{corner} \citep{ForemanMackey2016}.}
    \label{fig:corner}
\end{figure}

We show the joint posterior distributions for $R_p/R_*$, $b$, and $\rho_*$
from the fit in Figure~\ref{fig:corner}. The fit yields an orbital period
of $P = 4.644$\,d and a planet radius of 7.2\,$R_\oplus$
(Table~\ref{tab:derived}). The impact parameter is only moderately
constrained ($b = 0.28^{+0.20}_{-0.18}$), a consequence of the
well-known correlation between $b$ and $\rho_*$ in transit fits, visible
as the elongated shape of the corresponding panel in
Figure~\ref{fig:corner}. Despite this, none of our posterior samples
correspond to a grazing transit configuration, so we conclude the
transit geometry is non-grazing. Full posteriors, including the
GP and instrumental nuisance parameters, are reported in
Table~\ref{tab:tab5}, with the derived transit and planetary
parameters summarized separately in Table~\ref{tab:derived}.

\input{table5_new.tex}

\input{table5b.tex}

In addition to this global joint fit, we performed a
depth-focused refit of each instrument individually, fixing the impact
parameter and limb-darkening coefficients to their values from the
joint fit above while allowing the transit depth and per-instrument
polynomial detrending coefficients to vary freely.  Since
band-dependent stellar activity could imprint on the fitted depth, we
use the resulting per-band depths in our multicolor false-positive
analysis (Section~\ref{sec:mcph}).

\subsection{Transit Timing Variations}
\label{sec:ttvs}

To search for transit timing variations (TTVs), we measured individual mid-transit 
times across both TESS light curves and ground-based follow-up photometry 
(Swope, SSO, CTIO, and McDonald). Partial transits were excluded from the ephemeris fit analysis but their
midpoints were recorded for a consistency check. For each epoch, the transit shape and band-dependent 
limb-darkening parameters were fixed to the median posterior values obtained from Section~\ref{sec:juliet}. 

We evaluated the transit timing behavior by fitting a linear ephemeris to the TESS transits to establish an initial baseline, and tested
its predictive accuracy against independent ground-based observations. We then performed a joint linear fit across the full $\sim$418-day 
baseline, alongside a quadratic fit to search for long-term variations. 

Our $O-C$ analysis yields no significant departures from a constant orbital period. The predicted 
ground-based transit times agree with the TESS-only ephemeris within $1\sigma$, and the quadratic ephemeris 
reveals no detectable curvature across the extended baseline.
We therefore find no evidence for transit timing variations in TIC~88297141.

\section{False Positive Assessment}
\label{sec:validation}

To validate the planet, we must demonstrate that the data are better
explained by a transiting planet than by an astrophysical false
positive.  We assess various categories of false-positive scenarios,
including unresolved background eclipsing binaries (BEB), hierarchical
eclipsing binaries (HEB), and the possibility that the transit is
caused by an eclipsing binary (EB).

\begin{figure*}[t]
    \centering
    \includegraphics[width=\textwidth, height=0.7\textheight, keepaspectratio]{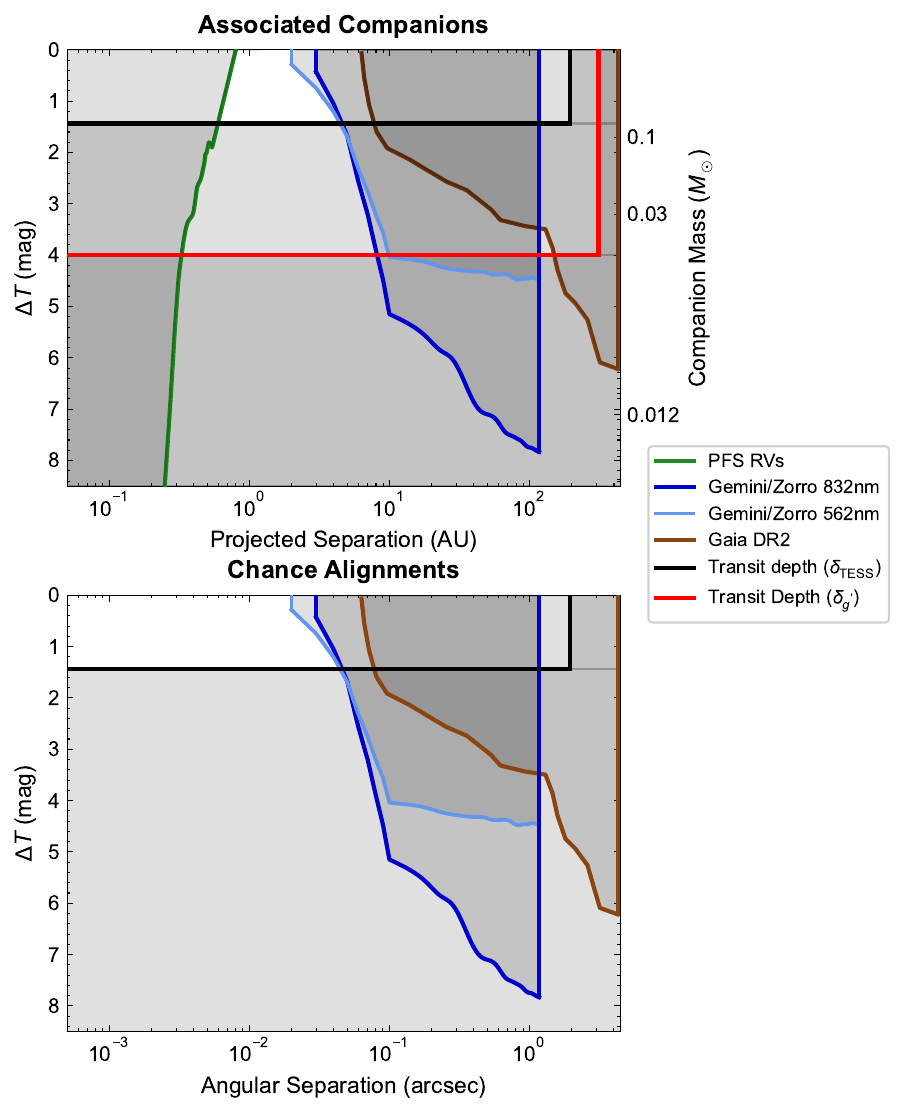}
    \caption{\textbf{Astrophysical false-positive scenarios.}
    \textit{Top}: For bound companions. The right-hand companion mass axis reflects the 10\,Myr mass--luminosity relation derived from MIST models \citep{Choi2016, Dotter2016} for $M > 0.1\,M_\odot$ and COND03 models \citep{Baraffe2003} for $M < 0.1\,M_\odot$ (see Section~\ref{sec:conversion}).
    \textit{Bottom}: For unassociated companions along the same line of sight (chance
    alignments). Gray regions are ruled out by at least one
    constraint.}
    \label{fig:fpplot}
\end{figure*}

Figure~\ref{fig:fpplot} summarizes these astrophysical false-positive
scenarios along with the exclusion zones established from our
follow-up data. In this section we describe each constraint in turn,
then present a calculation using \texttt{TRICERATOPS}
\citep{Giacalone2021} demonstrating that the probability of
TIC~88297141\,Ab being an astrophysical false positive is small enough
to validate it as a planet.

\subsection{Conversion to Mass Constraints}
\label{sec:conversion}

To convert our observational contrast limits into mass constraints for
potential bound companions, we used a composite mass-luminosity
relation following the methodology outlined by \citet{bouma2020}. For
the substellar regime ($M < 0.1\,M_\odot$), we adopted the COND03
evolutionary models \citep{Baraffe2003}, while for the stellar regime
($M > 0.1\,M_\odot$), we used the MESA Isochrones and Stellar
Tracks \citep[MIST;][]{Choi2016, Dotter2016}.  Although the estimated
age of our system is approximately 15\,Myr
(Section~\ref{sec:membership}), the available discrete model grids
necessitated a choice between the 10\,Myr and 30\,Myr isochrones. We
opted to evaluate the models at 10\,Myr to approximate the companion's
state of early contraction.  To convert the theoretical effective
temperatures and bolometric luminosities into expected instrumental
magnitudes, we made the simplifying assumption that all sources emit
as blackbodies.  By integrating these theoretical blackbody spectra
over the measured transmission functions for the TESS, Gemini/Zorro
562\,nm, and Gemini/Zorro 832\,nm bandpasses, we calculated the
expected magnitudes of hypothetical companions across our mass grid.
This approach allowed us to construct a continuous mapping from our
derived $\Delta m$ thresholds to physical companion mass limits.

\subsection{Constraints on False Positive Scenarios}
\label{sec:fpscenarios}

\subsubsection{Transit Depth}
\label{sec:trnsdpth}

To quantify the brightness limits on potential undetected stellar
contaminants or background eclipsing binaries (BEBs), we use the
geometric constraints outlined by \citet{Seager2003}.  In a central
transit scenario ($b=0$), the maximum intrinsic flux deficit $\Delta
F_{\text{real,max}}$ produced by a fully eclipsing companion is
bounded by the ingress ratio:

\begin{equation}
    \Delta F_{\text{real,max}} = \left(\frac{T_{12}}{T_{13}}\right)^2
\end{equation}

where $T_{12}$ represents the ingress duration $(T_{14} - T_{23})/2$
and $T_{13}$ is the duration from first to third contact $(T_{14} +
T_{23})/2$. 

The maximum allowable flux dilution factor between the unblended
maximum transit depth and the observed depth $\delta_{\text{obs}} =
(R_p / R_*)^2$ translates into a magnitude difference constraint
$\Delta m$:

\begin{equation}
    \Delta m \le 2.5 \log_{10}\left( \frac{(T_{12}/T_{13})^2}{\delta_{\text{obs}}} \right)
\end{equation}

Propagating our posterior parameter distributions through this
relation yields a $99.9\%$ upper bound of $\Delta m = 1.435$, with the
corresponding mass lower bound of $0.13\,M_\odot$. For TIC~88297141A
($T=13.1$), any blended star invoked to explain the transit depth must
be brighter than $T=14.5$.  We thus significantly constrain the viable
false-positive parameter space.

\subsubsection{Gaia}
\label{sec:gaiafp}

The \textit{Gaia} curve in Figure~\ref{fig:fpplot} combines Gaia's
ability to resolve companions directly as separate point sources at
wide separations with its sensitivity to unresolved companions via
excess noise in the astrometric fit at close separations. The curve is
interpolated from Figure 4 of \citet{rizzuto2018}. TIC~88297141 has a
RUWE statistic of 1.138, consistent with a single-star astrometric
solution and indicating no additional, unresolved companion within the
separations probed by this diagnostic. As highlighted before, the
system does host a wide, comoving companion (TIC~1450801333;
Section~\ref{sec:companion}), with a Gaia DR3 parallax and
proper motion consistent with gravitationally bound pair. At this 
separation, the interpolated Gaia curve indicates a detection limit of 
$\Delta G \approx 5.3$ mag: much fainter than the companion's observed 
$\Delta G = 2.13$ mag. Consequently, its presence is expected on imaging 
grounds and does not impact the RUWE-based limit above.

\subsubsection{Gemini Speckle Imaging}
\label{sec:gemfp}

The contrast limits obtained through Gemini/Zorro speckle imaging at
562 nm and 832 nm (Section~\ref{sec:speckle}) are shown in
Figure~\ref{fig:fpplot}. No additional companions were detected within
the instrument's field of view; the known companion lies outside the
Zorro field of view and is not constrained by this curve. Native
contrasts were converted to companion mass using the mass-magnitude
relation described in Section~\ref{sec:conversion} for the Zorro
bandpasses, then re-expressed as $\Delta T$ using MIST synthetic
photometry for stellar-mass companions and a blackbody approximation
for substellar-mass companions, for direct comparison with the other
panels in Figure~\ref{fig:fpplot}.

\subsubsection{Radial Velocity Constraints}
\label{sec:pfsfp}

We used the PFS radial velocities described in Section~\ref{sec:rv} to
test for both the transiting planet's own reflex motion and the
possible presence of additional, undetected companions in the system.
With only four independent epochs spanning a $\sim$43 day baseline,
however, a fully unconstrained Keplerian fit is poorly determined, and
we adopt a tiered approach of three models of decreasing complexity. 

In Model 1, we fix the orbital period to the photometrically
determined ephemeris and fit a Keplerian, testing directly whether the
data are consistent with the planet's own Doppler signal. In Model 2,
we instead let the period vary freely, to search more broadly for any
periodic signal present in the data; as discussed below, this fit does
not converge given the sparse sampling. Motivated by that
non-convergence, Model 3 takes a simpler, purely analytic approach:
fitting a linear trend to place an upper limit on any long-term
radial-velocity drift, which we translate into limits on the mass and
separation of a hypothetical outer companion under the assumption of a
circular orbit, and whose contrast curve is plotted in
Figure~\ref{fig:fpplot}.

\paragraph{Model 1: Fixed Period}

We first fixed the orbital period and $t_{0}$ to the transit-derived
ephemeris ($P_{\rm orb}=4.64$ d), assuming a circular orbit ($e=0$).
The free parameters for this fit were the RV semi-amplitude $K$, the
systemic velocity $\mu_{\rm PFS}$, and a log-uniform jitter term
$\sigma_{\rm w,PFS}$ to account for excess scatter beyond the formal
RV uncertainties. We fit this model using \texttt{juliet}, which calls
\texttt{radvel} \citep{Fulton2018} to construct the Keplerian RV
model. To test the dependence of our results on the adopted prior, we
performed two fits: one with a physical, non-negative prior on the RV
semi-amplitude ($K\sim\mathcal{U}[0,3500]$ m\,s$^{-1}$), and a second
with a symmetric prior ($K\sim\mathcal{U}[-3500,3500]$ m\,s$^{-1}$)
that removes the hard boundary at $K=0$. This ceiling corresponds to
$M_{\mathrm{p}} = 13\,M_{\rm J}$, the conventional deuterium-burning
boundary between planetary and brown-dwarf companions
\citep{Spiegel2011}, chosen independent of the RV data themselves. We
confirmed this choice does not truncate the posterior: with a narrower
ceiling of 1000\,m\,s$^{-1}$ ($\sim$3.7\,$M_{\rm J}$), the 3$\sigma$
tail of the $K$ posterior sits at the prior edge (996\,m\,s$^{-1}$), a
clear truncation artifact, whereas under the adopted 3500\,m\,s$^{-1}$
ceiling the posterior converges well within the prior support
(3$\sigma$ tail at $\sim$2230\,m\,s$^{-1}$).

Rather than assessing the significance of the RV signal from the
posterior distribution of $K$ alone, we adopt a Bayesian model
comparison approach. Bayesian evidence provides a statistically
rigorous measure of whether the increased complexity of a Keplerian
model is warranted by the data, naturally penalizing unnecessary model
complexity through the marginal likelihood
\citep{Feroz2011,Nelson2020}. Following the implementation adopted in
\texttt{juliet}, we compare the Bayesian evidence of each Keplerian
model against a null model in which the RV semi-amplitude is fixed to
$K=0$. The strength of evidence is quantified by

\begin{equation}
\Delta\ln Z = \ln Z_{\rm Kep} - \ln Z_{\rm flat},
\end{equation}

where positive values favor the Keplerian model and negative values
favor the null model. We interpret $\Delta\ln Z$ using the empirical
Jeffreys scale as summarized by \citet{Trotta2008} and commonly
adopted in Bayesian exoplanet analyses \citep{Feroz2011}, for which
$|\Delta\ln Z|<1$ is inconclusive, $1\leq|\Delta\ln Z|<2.5$
corresponds to weak evidence, $2.5\leq|\Delta\ln Z|<5$ to moderate
evidence, and $|\Delta\ln Z|\geq5$ to strong evidence.

For the flat model we obtain $\ln Z_{\rm flat}=-32.37$. The Keplerian
models yield $\ln Z_{K\ge0}=-32.84$ and $\ln
Z_{K\in[-3500,3500]}=-33.09$, corresponding to $\Delta\ln Z=-0.48$ and
$-0.72$, respectively. Under the Jeffreys/Trotta interpretation, both
comparisons are inconclusive, and their agreement across the two $K$
priors shows this is not an artifact of the non-negativity constraint.
We therefore find no Bayesian evidence for a Keplerian signal at the
known transiting planet period.

Despite the inconclusive evidence comparison, we use the resulting $K$
posterior to place an upper limit on the mass of any possible
companion locked to the orbital period.  We find $M_{\mathrm{p}}\sin i
< 4.7\,M_{\rm J}$ at $2\sigma$ and $< 7.9\,M_{\rm J}$ at $3\sigma$
(symmetric prior).  This limit is stable across
prior widths once the posterior is no longer truncated.

\paragraph{Model 2: Free Period}

Because a fixed-period fit alone cannot exclude a companion at a
different period, we also let the orbital period vary freely in a
circular Keplerian fit (\texttt{radvel}; \citealt{Fulton2018}, via
\texttt{juliet}), to map the region of $(K, P)$ space disfavored by
the data. With only four epochs over a $\sim$43 day baseline, this fit
does not converge to a well-defined solution.  Considered on their
own, the period and semi-amplitude marginals appear reasonably peaked:
$\ln P_b = 2.45^{+0.16}_{-0.43}$ ($P \approx 11.6^{+2.0}_{-4.0}$ days)
and $\ln K_b = 6.37^{+0.39}_{-0.54}$ ($K \approx
585^{+278}_{-245}$\,m\,s$^{-1}$). The time of conjunction, however,
$T_{\rm conj,b} = 2460830.92^{+350.33}_{-297.27}$ BJD, is essentially
unconstrained. Its uncertainty exceeds the period itself by more than
an order of magnitude, indicating that this apparent structure
reflects projection of a highly degenerate, multimodal posterior into
one dimension rather than a uniquely identified orbit.

As a robustness check, we reran the fit with the period prior's lower
bound raised above 11.6 days, explicitly excluding this peak, to test
whether the sampler would find an independent alternative. Instead,
the period posterior goes essentially flat over the remaining range,
nominally preferring an unphysical $\sim 1.5 \times 10^{8}$ day period
with an even larger uncertainty. The semi-amplitude posterior
similarly broadens, becoming consistent with zero, while the jitter
term grows substantially ($\sim$95 to $\sim$329\,m\,s$^{-1}$) to
absorb the scatter that the excluded solution had previously
explained. Removing the dominant local peak therefore eliminates any
period preference, rather than revealing a competing one. This is
further evidence that our four sparse epochs do not carry enough
information to constrain a free-period Keplerian model.

\paragraph{Model 3: An Analytical Approach}

Given the poor convergence of the free-period fit (Model 2), we
additionally fit a weighted linear trend to the four PFS epochs to
place an analytic, non-parametric bound on any long-term
radial-velocity drift. The raw formal fit yields a slope of $-4.958
\pm 0.895$\,m\,s$^{-1}$\,day$^{-1}$, but this line fit shows a large
reduced $\chi^2$ ($\chi^2/{\rm dof} = 65.1$). To check whether this
poor fit reflects genuine excess scatter rather than an inadequate
functional form, we also fit a flat (no-trend) model to the same
epochs; if a real linear trend were driving the residuals, the line
fit's reduced $\chi^2$ should improve relative to the flat case.
Instead, the flat fit shows a comparably large reduced $\chi^2$
($\chi^2/{\rm dof} = 53.6$), indicating that the poor fit is not an
artifact of assuming a linear trend, but reflects real point-to-point
scatter roughly $7$--$8\times$ larger than the formal per-visit errors
predict. This mirrors the procedure used in Section~\ref{sec:rv} when
binning the three exposures per night into a single epoch, where we
similarly rescaled the formal uncertainties by the square root of the
reduced $\chi^2$ to account for excess intra-night scatter. We
therefore apply the same rescaling here, multiplying the formal slope
uncertainty by a factor of 8.07, which gives a slope of $-4.96 \pm
7.22$\,m\,s$^{-1}$\,day$^{-1}$ and a 3$\sigma$ upper limit of $|dv/dt|
< 26.62$\,m\,s$^{-1}$\,day$^{-1}$, corresponding to a total $\Delta$RV
bound of 1.138\,km\,s$^{-1}$ over the $\sim$43 day baseline. 

To translate this $\Delta$RV bound into companion limits, we computed,
over a grid of semi-major axis $a$ and companion mass $m_2$, the
circular-orbit RV semi-amplitude
 
\begin{equation}
    K = \left(\frac{2\pi G}{P}\right)^{1/3} \frac{m_2}{(M_1+m_2)^{2/3}},
\end{equation}
 
adopting $M_1 = 0.30\,M_\odot$. Because the 43-day baseline may sample
only part of a longer-period orbit, we excluded a companion only if
its minimum possible RV change over the baseline, $\Delta v_{\rm min}
= K\left[1 - \cos\left(\pi\,\Delta T/P\right)\right]$, corresponding
to the least favorable orbital phase, exceeded the measured bound.  We
converted the resulting exclusion envelope to $\Delta m$ via the
mass-luminosity mapping of Section~\ref{sec:conversion}, and show it
in Figure~\ref{fig:fpplot}.

\subsubsection{Multicolor Photometry and HEB Scenarios}
\label{sec:mcph}

The most plausible HEB scenarios for TIC~88297141 involve a pair of
eclipsing low-mass companions orbiting the primary. We consider two
mass regimes: a stellar regime spanning the hydrogen-burning limit
($\sim0.075\,M_\odot$) up to the primary's own mass
($\sim0.30\,M_\odot$), and a substellar regime spanning the lowest
tabulated mass in the Baraffe/COND evolutionary models
($\sim0.5\,M_{\rm J}$) up to the primary mass. We adopt this upper
bound because a more massive, and hence brighter, unresolved stellar
pair would be expected to produce a distorted spectral energy
distribution, spectroscopic line contamination, or an elevated RUWE,
none of which are observed (Section~\ref{sec:gaia},
Section~\ref{sec:sed}). \texttt{TRICERATOPS} also evaluates this
configuration but fits only the TESS
light curve, with companion flux ratios drawn from theoretical
mass-luminosity relations rather than measured color (Section~\ref{sec:triceratops}). Our analysis
below instead places a constraint directly from the observed color
dependence of the transit depth.

Because TIC~88297141 is itself a mid-to-late M dwarf ($T_{\rm
eff}=3168$~K), the color contrast between the target and a putative
eclipsing pair drawn from either mass range is  weaker than in the
case of a hotter primary star. Nonetheless, at a fixed mass, a cooler
eclipsing pair still produces a somewhat redder eclipse than a transit
of the primary, so color-dependent depth limits retain some diagnostic
power, strongest in the bluest available bandpass.  We therefore
restrict this analysis to our CTIO SDSS $g'$ light curve.  We fit our
light curve individually using machinery described in
Section~\ref{sec:juliet}, fixing the impact parameter and
limb-darkening coefficients to their global joint-fit values and
allowing the transit depth and a local quadratic trend to vary freely,
with the ephemeris fixed from the TESS-only fit
(Section~\ref{sec:tess}). The resulting $2\sigma$ lower limit on the
transit depth was $4.32$~ppt.

Subsequently, we translated this limit into a mass exclusion,
interpolating stellar parameters from MIST isochrones and substellar
parameters from Baraffe/COND models, both at 10~Myr, and assuming
blackbody spectral energy distributions with the SVO Filter Profile
Service transmission curve for $g'$. Scanning the stellar regime ($M_2
= 0.075$--$0.30\,M_\odot$, mass ratios $M_3/M_2 = 0.1$--$1$) yields no
excluded masses: the maximal predicted depth, produced by
near-equal-mass (twin) tertiaries, exceeds the observed $2\sigma$
floor even at the hydrogen-burning limit.

In the substellar regime, we considered the limiting case of two
equal-mass companions in a perfectly central, total mutual eclipse --
the configuration that maximizes eclipse depth for a given total
system mass. Substellar parameters were interpolated from the
Baraffe/COND models down to the grid's lowest tabulated mass
($\sim0.0005\,M_\odot$, $\sim0.5\,M_{\rm J}$). We find that this
maximal equal-mass configuration reproduces the observed $g'$-band
depth only for $M_2 \gtrsim 0.02\,M_\odot$, corresponding to a
magnitude difference of $4.003$; below this mass, even the deepest
possible eclipse falls short of the observed $2\sigma$ floor,
excluding the substellar HEB scenario for $M_2 \lesssim
0.02\,M_\odot$. We further note that a genuinely eclipsing equal-mass
substellar pair, at the period and inclination required to alias into
our detected ephemeris, is itself an a priori unlikely configuration
given the low observed binary fraction among brown dwarfs; we do not
attempt to quantify this probability here, and rely on the depth
argument alone.

The substellar regime constraint is shown in red in
Figure~\ref{fig:fpplot}. The constraint is comparatively weak as it excludes only the narrow
substellar corner $M_2\lesssim0.02\,M_\odot$. The weak constraint is a
direct consequence of the primary's own cool temperature diminishing
the available color contrast; the achromatic geometric argument of
Section~\ref{sec:trnsdpth} places a stronger bound on any diluting
companion, independent of its color.

\subsection{False Positive Probability}
\label{sec:triceratops}

We explored the likelihood of false positive scenarios using
\texttt{TRICERATOPS}, a Bayesian tool that computes the false positive
probability (FPP) and nearby false positive probability (NFPP) for a
given transit signal. For each scenario, \texttt{TRICERATOPS}
evaluates the likelihood that the observed transit shape is consistent
with an astrophysical false positive configuration,  such as a
background eclipsing binary (BEB) or a nearby eclipsing binary (NEB),
relative to the planet hypothesis. Using the MAST module of
\texttt{astroquery} \citep{ginsburg2019}, the tool first queries the
TESS Input Catalog (TIC), which is itself built primarily from
\textit{Gaia} astrometry and photometry, for all stars within a radius
of 10 pixels of the target.  These known resolved sources build the
likelihoods for the NEB-type scenarios.  Background stellar
populations are drawn from TRILEGAL \citep{Girardi2005} simulations of
the Galactic field to build the likelihood of background eclipsing
binary scenarios caused by potential unresolved stars. The contrast
curve from Gemini-South speckle imaging was passed directly to
\texttt{TRICERATOPS} to constrain the parameter space of viable
configurations. Because the SPOC pipeline's CROWDSAP metric, i.e., the
fraction of flux within the photometric aperture attributed to the
target star, as estimated from the pipeline's modeled point-response
function and nearby catalog sources, differs between sectors
(CROWDSAP~$=0.364$ for Sector~92 and $=0.619$ for Sector~91) and
cannot be applied consistently to a single stitched light curve, we
ran \texttt{TRICERATOPS} independently on the PDCSAP light curves from
Sector~91 and Sector~92. The differing CROWDSAP values reflect SPOC's selection of a larger, 
five-pixel cross-shaped optimal aperture for Sector~92, versus a single pixel centered on the
target for Sector~91. To corroborate our split approach, we refit
the transit model (Section~\ref{sec:juliet}) to 
each sector's light curve independently, recovering depths
consistent within $1\sigma$.

For every star within its search radius, \texttt{TRICERATOPS} models
each star's point-spread function as a circular Gaussian of fixed
width ($\sigma=0.75$ pixels, held constant across the detector;
\citealt{Giacalone2021}), and computes a raw flux ratio $X_{i,\rm
raw}$ for each star from its PRF-weighted, aperture-integrated flux
relative to the summed flux of every star in the field.
\texttt{TRICERATOPS} then converts an observed transit depth into the
physical eclipse depth required on each candidate star via
$\delta_{s,i} = \delta_{\rm obs}/X_i$.

Because the CROWDSAP metric already rescales the PDCSAP flux to remove
the target's own aperture contamination, any potential signal
originating from a contaminating star is correspondingly amplified
relative to its true, diluted appearance in the raw SAP data.  Thus,
if \texttt{TRICERATOPS}'s default flux ratios are left unscaled, this
amplification causes the tool to infer an unphysically large eclipse
depth ($>100\%$) for these scenarios, incorrectly discarding viable
false-positive scenarios.  To correct for this, we rescaled every flux
ratio returned by \texttt{TRICERATOPS} within the search radius,
including the target's own, by $1/\rm CROWDSAP$, using each sector's
own value:
\begin{equation}
    X_{i,\,\rm eff} = \frac{X_{i,\,\rm raw}}{\rm CROWDSAP}.
\end{equation}
Since this factor is common to every star in a sector's aperture, the
relative weighting between candidates is preserved and only the
overall normalization is corrected.

Additionally, stars lacking TIC-reported parameters default to solar
values, which, for M dwarfs, overstates the viable eclipse-depth
range. We substituted directly derived parameters for both the primary
and the bound companion in place of these defaults. For the primary,
we adopted the radius obtained from our SED fit
(Section~\ref{sec:sed}) and a mass derived from that same radius
combined with the median stellar density recovered from our posterior,
$M=4/3\,\pi R^3 \bar\rho$.

The companion likewise has no TIC-reported mass or radius. Given its
TIC-reported effective temperature ($T_{\rm eff}=3016$~K), consistent
with an M dwarf, we adopt an approximate mass and radius from the
\citet{PecautMamajek2013} tables as a more physically reasonable
substitute for the solar default, rather than pursue a full
spectroscopic or SED characterization of this star.
These parameters improve on the solar defaults but are not a precise
characterization, since the companion is itself pre-main-sequence.

Updating \texttt{TRICERATOPS} with these primary and companion
parameters in place of default/catalog values reduces NFPP to 0.00087
for Sector~92 and 0.001244 for Sector~91, and we recovered $\rm FPP
\approx NFPP$ for both sectors, indicating that essentially all
false-positive probability at this stage is attributable to off-target
(nearby-star) scenarios rather than to any on-target configuration,
prior to the direct positional exclusion described below. 

For our final adopted false positive probabilities, we 
include constraints from our ground-based seeing-limited photometry.
Excluding the target, the closest star in this scenario
space is the bound companion at 
$2.568''$; the next closest star lies at
$3.927''$. Our observations from McDonald ($1.17''$
aperture; SDSS $i'$, transit ingress), Las Campanas 
($1.305''$ aperture; Sloan $r'$, egress), and Siding Spring
($1.945''$ aperture; SDSS $i'$, full transit;
Table~\ref{tab:groundphot}) all employ apertures narrower than the
$2.568''$ separation to the bound companion, and each detects the transit
ingress at the predicted time and depth.
These detections therefore exclude 
every neighboring star, including the bound
companion, as the source of the signal.  As a further check, 
where seeing permitted, we extracted light curves from smaller
apertures centered on the target (A), the bound companion (B), and
at a control position opposite B;  this exercise localized the transit
to the target star.  Following the convention of
\citet{Giacalone2021}, which fixes NFPP to zero once on-target origin
is confirmed by direct positional constraints, we adopt $\rm NFPP=0$
for both sectors. This leaves a total $\rm FPP\approx3\times10^{-6}$
for Sector~92 and $\rm FPP\approx1\times10^{-5}$ for Sector~91, in
both cases attributable to residual on-target (non-nearby-star)
scenarios. TIC~88297141\,Ab therefore satisfies the statistical
validation criteria for a bona fide planet.

\section{Discussion}
\label{sec:discussion}

\subsection{TIC~88297141\,Ab in Context}
\label{sec:context}

\begin{figure}
    \centering
    \begin{minipage}{0.48\textwidth}
        \centering
        \includegraphics[width=\linewidth]{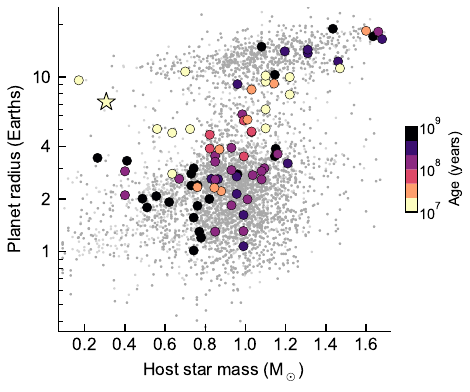}
    \end{minipage}
    \begin{minipage}{0.48\textwidth}
        \centering
        \includegraphics[width=\linewidth]{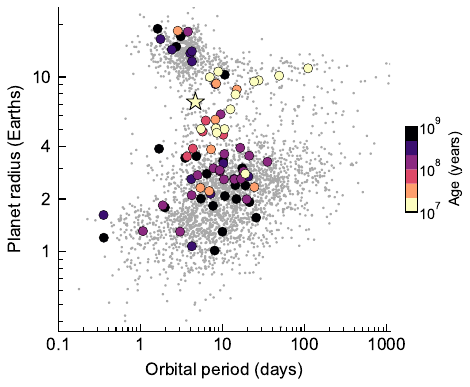}
    \end{minipage}
    \caption{\textbf{TIC~88297141\,Ab compared to known transiting
    planets.} Planets with ages between $10^{7}$ and
    $10^{9}\,\mathrm{yr}$ are color-coded by age; the sample is selected 
    using a $2\sigma$ upper age limit of $t + 2\sigma_t < 1\,Gyr$.  
    \textit{Top:} Planet radii versus host stellar mass. The ages and radii 
    are from the NASA Exoplanet Archive dated 2026 Aug 12. TIC~88297141\,Ab 
    is emphasized with a star.  \textit{Bottom:} Planet radii versus
    orbital periods. TIC~88297141\,Ab is emphasized with a star.}
    \label{fig:sizeage}
\end{figure}

Having validated TIC~88297141\,Ab's planetary nature, here we discuss
its properties in the context of the known population of young
transiting planets.

TIC~88297141\,Ab adds to a small Sco-Cen transiting sample whose
members are individually remarkable.  HIP~67522\,b, announced as a
17\,Myr hot Jupiter \citep{rizzuto2020}, proved to have a mass of only
$\sim$14\,$M_\oplus$ from JWST transmission spectroscopy
\citep{thao2024b}, and has since been joined by a second,
near-resonant giant \citep{barber2024a}.
TOI-1227\,b pairs a 9.5\,$R_\oplus$ radius with a 27\,day orbit around
a $\sim$0.17\,$M_\odot$ star, with an RV upper limit that likewise
implies a low density \citep{mann2022} and transit-timing variations
suggestive of a second planet \citep{almenara2024}.  Most recently,
HD~114082\,b and c ($P=226$ and $\sim$314\,days) are both
Jovian-radius yet low-density, orbiting within a two-component debris
disk \citep{zakhozhay2022, delburgo2026}.  Where mass constraints
exist for these young giants, they consistently indicate low
densities.

Figure~\ref{fig:sizeage} shows the planet relative to the broader
population of young transiting planets.  The top panel plots the planet
radius--host stellar mass plane, following the comparison
from \citet{Mann2017}. Because higher-mass stars host more
massive protoplanetary disks \citep{Andrews2013, Pascucci2016} and are
thus statistically more likely to form large planets
\citep{Johnson2010, Mulders2015}, sub-Saturn-sized planets might be
expected to become intrinsically rarer for lower-mass stars.
Consistent with these findings, mature-population
M-dwarf occurrence studies from Kepler, K2, and TESS
\citep{dressing2015, Ment2023, ullman2025, 2026Gillis} consistently find few or
no Neptune-to-Saturn-sized planets: the deepest TESS survey of
mid-to-late M dwarfs to date \citep{2026Gillis} recovers super-Earths
outnumbering sub-Neptunes 5.5:1 and finds no sub-Saturns at all. However, such
surveys target main-sequence stars and do not constrain occurrence as a
function of age. 

TIC~88297141\,Ab's youth may speak to this distinction: its radius
situates it within a population of $5$--$10\,R_\oplus$ planets
recently discovered around stars younger than 100 Myr, understood to
be the direct progenitors of the super-Earths/sub-Neptunes found
around Gyr-old stars \citep{Owen2020}. While some young
Neptune-sized planets are known around early-to-mid M dwarfs, such as
AU Mic b (22 Myr, 0.5\,$M_\odot$, 
4\,$R_\oplus$; \citealt{plavchan2020, Gilbert2022}),
sub-Saturn-sized planets around young, low-mass M dwarfs remain
rare, with TOI-1227b \citep{mann2022} ($\sim$11~Myr,
$\sim$0.17\,$M_\odot$, $\sim$9.5\,$R_\oplus$) standing as the only
comparable system.  Under the progenitor
picture above, its size best reflects residual formation heat not yet
radiated away, suggesting that TIC~88297141\,Ab is a still-contracting
progenitor of the smaller planets that dominate the mature
low-mass-star population. 

The bottom panel of Figure~\ref{fig:sizeage} situates the planet
within the youngest end of known transiting planets
directly.  Of the 82 transiting planets younger than 1~Gyr with robust
ages in the NASA Exoplanet Archive (2026 Aug 12), 13 are
``infants'' ($\log t < 7.33$), and eight of these fall within the
sub-Saturn radius range ($5$--$10\,R_\oplus$) highlighted in
Section~\ref{sec:intro} as sparsely sampled; TIC~88297141\,Ab is
the ninth. While the raw number of confirmed discoveries in this regime remains small, 
occurrence rate studies tailored to young TESS stars demonstrate that similarly sized planets 
($1.8$--$10\,R_\oplus$) are detectable and intrinsically more prevalent at young 
ages \citep{Vach2024, Fernandes2025}. TIC~88297141\,Ab therefore occupies an underrepresented territory in age--radius space, rather than an 
astrophysically unlikely region.

\subsection{Possible Futures}
\label{sec:futures}

\begin{figure}
    \centering
		\includegraphics[width=\linewidth]{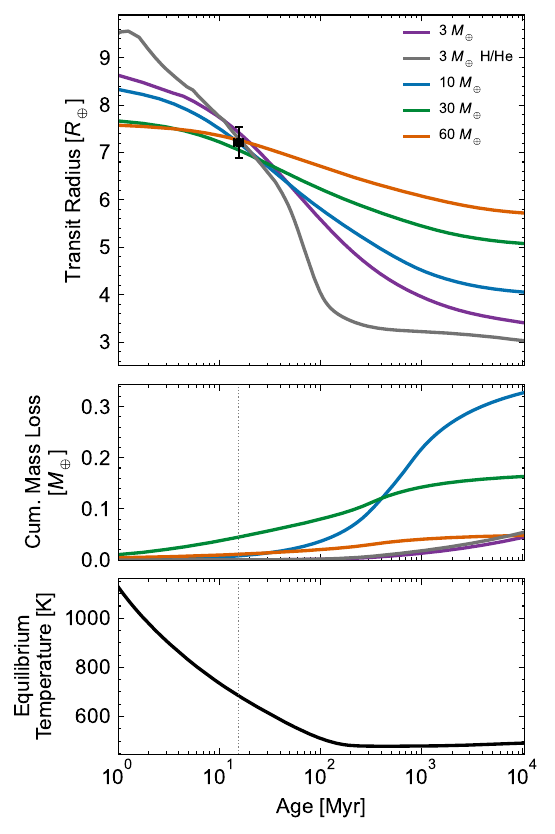}
    \caption{\textbf{Possible futures of TIC~88297141\,Ab.}
    \textit{Top:} Example radius evolution of \texttt{ORCHARD}
    tracks consistent with the present radius and age (black marker) across a 3--60 M$_\oplus$ mass range.
    Gray model is a pure H/He envelope-core model, while the rest have an envelope $Z=0.1$.
    \textit{Middle:} Cumulative mass lost to XUV photoevaporation; the dotted line marks the system age.
    \textit{Bottom:} Equilibrium temperature evolution according to MIST stellar evolution models of a 0.31 M$_\odot$ M-dwarf.}
    \label{fig:futures}
\end{figure}

We explored possible futures of TIC\,88297141\,Ab by computing a grid of
planetary evolution models using \texttt{ORCHARD}
\citep{TejadaArevalo2026}, a publicly available Henyey-evolution
code descended from \texttt{APPLE} \citep{Sur2024}.  The grid
comprised over 1,100 tracks spanning total masses of
3--80\,$M_\oplus$, envelope metal mass fractions of $Z \approx
0$--0.3, compact core mass fractions of up to $\sim0.9$, and
initial entropies ranging from cold to hot starts.  The interior
models are coupled to the non-gray radiative-convective atmospheric
boundary condition of \citet{Ohno2023}, which are metallicity and 
stellar-flux dependent, with an
evolving planetary equilibrium temperature set by the 
luminosity evolution of the 0.31\,$M_\odot$ host from MIST 
\citep{Choi2016, Dotter2016}.  For mass loss, we follow an XUV-driven
photoevaporation prescription implemented in \citet{Chen2016}, 
whose methodology is based on \citet{Ribas2005} and \citet{MurrayClay2009}.
We take the minimum of the energy-limited and recombination-limited 
rates, with heating efficiencies of 0.1--0.5 \citep{Owen2012} and 
the stellar XUV history parametrized following \citet{Jackson2012} and
\citet{Rogers2021}.  Core-powered and boil-off mass loss
\citep{ginzburg2018, owenwu2016, Misener2026} are not included; both could
plausibly accelerate the evolution described below.

Our models indicate that a wide range of masses and interior
compositions can yield the planet's present-day radius: tracks
spanning 3--80\,$M_\oplus$ reproduce $R_{\rm p} = 7.21 \pm
0.32\,R_\oplus$ at the system age through compensating combinations of
core mass, envelope metallicity, and initial entropy.  Two requirements nonetheless hold.  First, metal-poor, core-free
models overshoot the observed radius and are strongly disfavored:
acceptable fits require either envelope enrichment ($Z \gtrsim 0.1$)
or core mass fractions of $\approx$0.65--0.85.  Second, hot initial
entropies are disfavored, as they also overshoot the observed radius.
Meanwhile, photoevaporation plays only a minor role: the wind is
recombination-limited at early times, and by 15.6\,Myr the planet has
intercepted only $\approx$2\% of its lifetime XUV fluence and shed
$<$1\% of the mass it will ultimately lose.  Tracks with and without
mass loss are therefore indistinguishable today, and since
Kelvin--Helmholtz contraction is ongoing, the current radius provides
little direct leverage on the final one.

The allowed futures diverge over gigayear timescales depending
on the planet's mass and composition, as example models show in
Figure~\ref{fig:futures}. The transit radii are calculated 
following the prescriptions of \cite{Guillot2010}. Models with $M_{\rm p} \lesssim
10\,M_\oplus$ and photoevaporative mass loss contract below
4\,$R_\oplus$, with H/He-envelope models at high core mass fractions
reaching $\approx$3\,$R_\oplus$ (see gray model in top panel of Figure~\ref{fig:futures}); models with 10--20\,$M_\oplus$ settle
at 4--5\,$R_\oplus$; and more massive, metal-rich models remain at
$\approx$5--7\,$R_\oplus$.  The latter outcome would be unusual: among
mature planets orbiting similar M~dwarfs, the 4--7\,$R_\oplus$ range
is nearly unpopulated (Figure~\ref{fig:sizeage}), with TOI-3884\,b a
rare exception \citep{Almenara2022}.  A young analog for the massive,
core-dominated scenario does exist in TOI-837\,b \citep{bouma2020,
barragan2024}.  If instead TIC\,88297141\,Ab is drawn from the same
population as the known mature M-dwarf planets, the principle of
mediocrity suggests $M_{\rm p} \lesssim 20\,M_\oplus$, in which case
the planet would be a Neptune, or even a sub-Neptune observed early in its contraction.
A dynamical mass measurement would test this assumption.

\subsection{Follow-up Opportunities}
\label{sec:fop}

While we have statistically validated
TIC~88297141\,Ab, it is not yet 
``confirmed'': as
Figure~\ref{fig:fpplot} shows, there remain regions of parameter space
where a background eclipsing binary (BEB) could reproduce the eclipse
depth. However, BEB (and HEB) configurations produce grazing, V-shaped
transits, whereas the transit of TIC~88297141\,Ab is U-shaped, which
is why our false positive probabilities strongly disfavor this scenario.

Still, confirming the planetary nature of TIC~88297141\,Ab could
proceed along two independent, complementary paths: a spectroscopic
detection via the Rossiter--McLaughlin effect, or an atmospheric
detection via transmission spectroscopy. An RM detection consistent
with the photometric signal would help rule out BEB and HEB scenarios
as it would establish that the eclipsing object is bound to the target
star. Combined with our mass upper limit derived from the PFS RVs,
these data would be enough to confirm that TIC~88297141\,Ab is a
planet. The maximum RM amplitude can be calculated following the
procedure of \citet{Gaudi2007},
\begin{equation}
\Delta V_{\mathrm{RM}} \approx f_{\mathrm{LD}} \cdot \delta \cdot
  v\sin i \cdot \sqrt{1-b^2} \approx 154 \sin i \ \mathrm{m\,s^{-1}}
\end{equation}
for
\begin{equation}
  f_{\mathrm{LD}} = 1 - u_1(1-\mu) - u_2(1-\mu)^{2}
\end{equation}
where $\mu \approx (1-b^{2})^{1/2}$, and $u_i$ denote the
limb-darkening parameters extracted from Table~\ref{tab:tab5}.  At
$T$=13.1, however, the star may stretch the limits of VLT
spectrographs like ESPRESSO;  the measurement may be more feasible
with future ELT-class spectrographs. Alternatively, 
transmission spectroscopy with JWST could confirm the planetary nature
of TIC~88297141\,Ab by detecting an atmosphere, while atmospheric
retrieval of the scale height could constrain the planet's
surface gravity and mass without requiring an RV or RM
detection at all. 

TIC~88297141 will be reobserved by TESS in Sectors 115 (2027 March
14--April 12) and 116 (2027 April 12--May 10), with Sector 116
observed by two cameras simultaneously. These data, expected to become
available circa July 2027, will extend our observational baseline from
441 days (Sector~91 through our latest ground-based transit on 2026
June~24) to 761 days, an increase of $\sim320$ days. This longer baseline will help further tighten the
transit ephemeris, which is crucial for potential RV or JWST follow-up. 
The additional sectors will also better constrain the transit shape, and consequently the planet
size, as well as improving our sensitivity to transit-timing
variations (TTVs) in search of other planets in the
system.

\vspace{0.5cm}
G.K. acknowledges support from the Carnegie Astrophysics Summer
Student Internship (CASSI) program, from the Caltech Summer
Undergraduate Research Fellowship (SURF) program, and from NASA grant 80NSSC25K0121 (G07140).
L.G.B. acknowledges support from the Carnegie Fellowship during the
project's early stages. 
This material is based upon work supported by the National Science Foundation Graduate Research Fellowship Program under Grant No. DGE-2439854. 
Any opinions, findings, and conclusions or recommendations expressed in this material are those of the authors and do not necessarily reflect the views of the National Science Foundation. 
A.W.B. thanks the LSST-DA Data Science Fellowship Program, which is funded by LSST-DA, the Brinson Foundation, the WoodNext Foundation, the Research Corporation for 
Science Advancement Foundation, and Kevin Wells; his participation in the program has benefited this work.
A.W.M. was supported by grants from the TESS Guest investigator program (80NSSC25K0113), 
the NSF CAREER program (AST-2143763), and NASA's Exoplanet Research Program (XRP 80NSSC25K7148).
K.A.C. acknowledges support from the TESS mission via subaward s3449 from MIT and NASA grants 80NSSC24K1889 and 80NSSC26K0081.
Funding for K.B. was provided by the European Union (ERC AdG SUBSTELLAR, GA 101054354). Funding for S.G.-G. was provided by the NASA NY Space Grant.
We thank the Magellan staff, including the telescope operators and
instrument specialists, for their support of the observations
presented in this work. We also thank Christopher R. Burns for reducing the Swope photometry presented in this work, using the CSPy pipeline, and Steven Giacalone for a helpful 
discussion on the application of \texttt{TRICERATOPS} \citep{Giacalone2021} 
to this system. This work used data from TESS \citep{ricker2015}.
We acknowledge the use of public TESS data from pipelines at the TESS Science Office and at the TESS Science Processing Operations Center.
Resources supporting this work were provided by the NASA High-End Computing (HEC) 
Program through the NASA Advanced Supercomputing (NAS) Division at Ames Research Center for the production of the SPOC data products.
The 2-minute cadence observations used in this work were proposed by
G07140 (PI: A.~Boyle). This research made use of the Montreal Open Clusters 
and Associations (MOCA) database, operated at the Montr\'eal Plan\'etarium
\citep{Gagne2026}. We used light curves from SPOC \citep{Jenkins2016}.
Funding for the TESS mission is provided by NASA's Science Mission
Directorate. This work is partly supported by JSPS KAKENHI Grant Numbers
JP24H00017, JP25K24620, JP26H01402, JP26K00755, and JP24K00689. 
This work makes use of observations from the LCOGT network. Part of the LCOGT telescope 
time was granted by NOIRLab through the Mid-Scale Innovations Program (MSIP). MSIP is funded by NSF.
This research has made use of the Exoplanet Follow-up Observation Program (ExoFOP; DOI: 10.26134/ExoFOP5) website, 
which is operated by the California Institute of Technology, under contract with the National Aeronautics and Space 
Administration under the Exoplanet Exploration Program.
Finally, this work also benefited from using the SVO Filter Profile
Service ``Carlos Rodrigo'', funded by
MCIN/AEI/10.13039/501100011033/ through grant PID2023-146210NB-I00.
\vspace{0.5cm}

{\it \large Contributions}: 
GK led the transit search, identified TIC~88297141\,Ab, led the
characterization and validation of the system, and wrote the
manuscript.
LGB advised GK, facilitated data acquisition, validated the transit
fits, and analyzed the Sco-Cen membership.
AWB designed the Sco-Cen target list and led G07140.
RTA computed and interpreted the planet evolution models.
Ground-based photometry was obtained by KB (SSO), JDL with NN, AF and TS
(CTIO Sinistro 1\,m), FPW with GH, VZ and SGG (CTIO and McDonald
Observatory), NM and LGB (Swope); with SOAR imaging from AWM and MGB.
The SSO, CTIO, and McDonald observations were coordinated by KAC.
DRC, SBH, CL, and SD obtained and reduced the Gemini/Zorro speckle
imaging.
PFS spectra were obtained by RPB, LGB, SV, MM, and the PFS team (JT, SAS,
JDC) and were reduced by RPB. DAC helped develop, operate, and validate the TESS SPOC data analysis.
All authors reviewed the manuscript.

\facilities{
  Swope,
  Magellan:Clay (PFS),
  SOAR:Goodman,
  Siding Spring Observatory:Sinistro,
  Cerro Tololo Inter-American Observatory:Sinistro,
  McDonald Observatory:Sinistro,
  Gemini South:Zorro,
  Gaia,
  TESS.
}

\software{
  w\=otan \citep{Hippke2019},
  TransitLeastSquares \citep{Hippke2019b},
  BANZAI \citep{McCully2018},
  AstroImageJ \citep{Collins2017},
  Prose \citep{prose},
  EAGLES \citep{jeffries2023},
  Comove \citep{Tofflemire2021},
  juliet \citep{Espinoza2019},
  batman \citep{batman},
  celerite \citep{ForemanMackey2017},
  dynesty \citep{Speagle2020},
  corner \citep{ForemanMackey2016},
  CSPy (\url{https://github.com/obscode/CSPy})
  TRICERATOPS \citep{Giacalone2021},
  RadVel \citep{Fulton2018},
  astroquery \citep{ginsburg2019},
  TRILEGAL \citep{Girardi2005},
  astroariadne \citep{Vines2022},
  astropy (\citealt{astropy2013,astropy2018,astropy2022}),
  matplotlib \citep{matplotlib},
  numpy \citep{numpy},
  pandas \citep{pandas},
  scipy \citep{scipy}.
}

\bibliography{bibliography}

\end{document}

%% file: vals.tex
\newcommand{\teff}{3168}

\newcommand{\period}{\ensuremath{4.644588_{-0.000026}^{+0.000026}}}

\newcommand{\fpp}{\ensuremath{1 \times 10^{-5}}}

\newcommand{\prot}{1.83}

%% file: apertures.tex
\begin{deluxetable*}{lcccccc}
    \tablecaption{Ground-Based Time-Series Photometry \label{tab:groundphot}}
    \tablehead{
    \colhead{Facility} & \colhead{Filter} & \colhead{UT Date} & \colhead{Coverage} & 
    \colhead{Aperture Radius} & \colhead{D} & \colhead{Cadence}
    }
    \startdata
    Swope 1\,m            & Sloan $r'$         & 2026 Mar 24 & Egress  & $1.305''$ & 0.995 & $\sim$89\,s \\
    CTIO (LCOGT 1\,m)      & SDSS $g'$+$i'$     & 2026 Apr 16 & Ingress & $3.112''$ &  0.920 & $\sim$49\,s \\
    McDonald (LCOGT 1\,m)  & SDSS $i'$          & 2026 Apr 16 & Ingress & $1.17''$  & 0.985 & $\sim$90\,s \\
    McDonald (LCOGT 1\,m)  & SDSS $g'$          & 2026 Apr 16 & Ingress & $5.07''$  & 0.881 & $\sim$127\,s \\
    SSO (LCOGT 1\,m)       & SDSS $i'$          & 2026 Apr 25 & Full    & $1.945''$ & 0.975 &  $\sim$90\,s \\
    CTIO 1\,m (non-TFOP)   & SDSS $g'$          & 2026 Jun 24 & Full    & $1.556''$ &  0.977 & $\sim$70\,s \\
    CTIO 1\,m (non-TFOP)   & PS $z_s$           & 2026 Jun 24 & Full    & $3.890''$ & 0.901 & $\sim$70\,s \\
    \enddata
\end{deluxetable*}

%% file: PFS_rv_table.tex
\begin{deluxetable}{ccccc}
    \tablecaption{PFS Radial Velocities of TIC\,88297141\label{tab:rv}}
    \tablehead{
    \colhead{BJD$_{\rm TDB}$} & \colhead{UT Date} & \colhead{N$_{\rm exp}$} & 
    \colhead{RV (m\,s$^{-1}$)} & \colhead{$\sigma_{\rm RV}$ (m\,s$^{-1}$)}
    }
    \startdata
    2461156.823 & 2026 Apr 26 & 3 & 117.74  & 45.969 \\
    2461157.838 & 2026 Apr 27 & 3 & 72.52   & 57.637 \\
    2461159.813 & 2026 Apr 29 & 3 & -527.75 & 40.328 \\
    2461199.581 & 2026 Jun 8  & 3 & -363.77 & 25.376 \\
    \enddata
    \end{deluxetable}

%% file: stellar_properties.tex
\newcommand{\Teff}{T_{\mathrm{eff}}}
\newcommand{\FeH}{[\mathrm{Fe/H}]}

\providecommand{\nolinenumbers}{}
\begin{table*}
\nolinenumbers
\centering
\footnotesize
\caption{\textbf{Literature and Measured Properties for TIC~88297141 and Its Bound Stellar Companion}}
\label{tab:stellar_params}
\begin{tabular}{@{}p{2.2cm}p{4.0cm}p{4.2cm}p{4.2cm}c@{}}
\toprule
\multicolumn{5}{c}{Other Identifiers} \\
\midrule
\multicolumn{5}{l}{Primary: TIC 88297141; Gaia DR3 4109766661272060288} \\
\multicolumn{5}{l}{Companion: TIC 1450801333; Gaia DR3 4109766661238033408; separation $2\farcs568$ NE ($\sim$260 AU) from primary} \\
\midrule
Parameter & Description & Primary (TIC 88297141) & Companion (TIC 1450801333) & Source \\
\midrule
\multicolumn{5}{l}{\emph{Astrometry}} \\
\addlinespace[2pt]
$\alpha_{J2016}$ & R.A.\ (hh:mm:ss) & 17:27:08.29 & 17:27:08.36 & 1 \\
$\delta_{J2016}$ & Decl.\ (dd:mm:ss) & -25:58:11.05 & -25:58:08.67 & 1 \\
$l_{J2016}$ & Galactic longitude \ (deg) &  0.3009 & 0.3016 & 1 \\
$b_{J2016}$ & Galactic latitude \ (deg) & +5.0509 & +5.0510 & 1 \\
\addlinespace[1pt]
$\pi$ & Gaia DR3 parallax (mas) & $10.0006 \pm 0.03$ & $10.0444 \pm 0.08$ & 1 \\
$d$ & Distance (pc) & $99.7$ (SED); $101$ (Gaia) & 99.4 (SED) & 1,2 \\
$\mu_\alpha$ & Proper motion in R.A.\ (mas~yr$^{-1}$) & \emph{$-11.1738 \pm 0.03 $} & \emph{$-11.3264 \pm 0.08 $} & 1 \\
$\mu_\delta$ & Proper motion in Decl.\ (mas~yr$^{-1}$) & \emph{$-36.5666 \pm 0.02 $} & \emph{$-33.7835 \pm 0.06$} & 1 \\ 
RV & Systemic RV (km~s$^{-1}$) & $-16.51 \pm 6.1$ & ---  & 1 \\
RUWE & Renormalized unit weight error & $1.138$ & $	1.090$ & 1 \\
$A_0$ & Gaia extinction estimate (mag) & $0.91$ & --- & 1 \\
\addlinespace[4pt]
\multicolumn{5}{l}{\emph{Photometry}} \\
\addlinespace[2pt]
$G$ & Gaia-$G$ mag & $14.289 \pm 0.003$ & $16.422 \pm  0.002$ & 1 \\
$Bp$ & Gaia-$Bp$ mag & $15.913 \pm 0.005$ & $18.433 \pm 0.071$ & 1 \\
$Rp$ & Gaia-$Rp$ mag & $13.035 \pm 0.006 $ & $14.946 \pm 0.005 $ & 1 \\
$T$ & TESS mag & $13.14$ & $15.14$ & 1,3 \\
$r$ & SkyMapper $r$-band mag & $15.112 \pm 0.006$ & --- & 4 \\
$i$ & SDSS $i$-band mag & $13.376 \pm 0.044 $ & --- & 5 \\
$g'$ & SOAR $g$-band mag & $16.267 \pm 0.019$ & $18.990 \pm 0.070$ & 6 \\
$r'$ & SOAR $r$-band mag & $15.114 \pm 0.014$ & $17.493 \pm 0.037$ & 6 \\
$i'$ & SOAR $i$-band mag & $13.549 \pm 0.014$ & $15.557 \pm 0.023$ & 6 \\
$g$ & Pan-STARRS $g$-band mag & $16.249 \pm 0.002$ & --- & 7 \\
$r$ & Pan-STARRS $r$-band mag & $15.109 \pm 0.006$ & $19.627 \pm 0.000 $ & 7 \\
$i$ & Pan-STARRS $i$-band mag & $13.572 \pm 0.001$ &$15.786 \pm 0.092 $& 7\\
$z$ & Pan-STARRS $z$-band mag & $13.085 \pm 0.023$ & $14.822 \pm 0.028 $ &7 \\
\addlinespace[4pt]
\multicolumn{5}{l}{\emph{Spectroscopic \& Derived Stellar Parameters}} \\
\addlinespace[2pt]
$\Teff$ & Effective temperature (K) & $3168^{+36}_{-71}$ & $2843^{+142}_{-183}$ & 3,8 \\
$\log g_*$ & Surface gravity (cgs) & $4.248^{+0.044}_{-0.046}$ & $4.644^{+0.224}_{-0.178}$ & 8\\
$\FeH^\dagger$ & Metallicity & $0.270$ &  $0.188$ & 8 \\
$v_{eq}$ & Equatorial rotation velocity (km~s$^{-1}$) & $20.36 \pm 0.85 $& --- (not measured) & 3,8 \\
Li EW & 6707.8~\AA\ EW (m\AA) & $345.3^{+36.1}_{-26.6}$ & --- (not measured) & 8 \\
$P_{\rm rot}$ & Rotation period (day) & $1.83 \pm 0.051 $  & --- (not measured) & 3 \\
Cluster Age & Isochrone age for local ensemble (Myr) & $15.6 \pm 1.6$ & assumed coeval & 9 \\
\multirow{2}{*}{Age}
& Isochrone (Myr)
& $15.3^{+3.2}_{-5.7}$
& ---
& 9 \\
& Lithium/EAGLES (Myr)
& $18.2^{+3.2}_{-15.7}$
& assumed coeval
& 10 \\
$R_*$ & Stellar radius ($R_\odot$) & $0.7375^{+0.0243}_{-0.0234}$ & $0.3944^{+0.1046}_{-0.0772}$ & 8 \\
\multirow{2}{*}{$M_*$}
& Empirical relation ($M_\odot$)
& $0.307^{+0.055}_{-0.072}$
& --- (not measured)
& 8 \\
& Dartmouth Magnetic Model ($M_\odot$)
& $0.352 \pm 0.028$
& $0.252^{+0.022}_{-0.012}$
& 11 \\
\bottomrule
\end{tabular}

\vspace{4pt}
\begin{minipage}{0.95\linewidth}
\footnotesize
\textbf{$\dagger$} Prior-dominated parameter. Median reported value is unconstrained by the available photometry (see text). \\ 
\textbf{Sources.} (1) Gaia DR3 \citep{GaiaDR32023}; (2) SED fit, this work; (3) TESS photometry \citep{Stassun2019}; (4) SkyMapper Survey \citep{SkyMapper4} (5) Sloan Digital Sky Survey \citep{Sloan} (6) SOAR photometry, A. Mann (priv.\ comm., this work); (7) Pan-STARRs Survey \citep{PS1} (8) SED fit (AstroARIADNE), this work; (9) \cite{Kerr2021} (10) Li~I EW + EAGLES v2, this work; (11) \citet{Feiden2016}
\end{minipage}
\end{table*}

%% file: table5_new.tex
\begin{table*}
\centering
\scriptsize
\caption{Priors and Posteriors for the Model Fitted to the TESS and Ground-based Data}
\label{tab:tab5}
\begin{tabular}{lllrrrrr}
\hline\hline
Param. & Unit & Prior & Median & Mean & Std. Dev. & 3\% & 97\% \\
\hline
\multicolumn{8}{l}{\textit{Sampled: physical}} \\
$P$ & day & $\mathcal{N}(4.64423; 0.01)$ & 4.644642 & 4.644643 & 2.05222e-05 & 4.644605 & 4.644682 \\
$t_0$ & day & $\mathcal{N}(3803.24; 0.1)$ & 3803.240 & 3803.240 & 0.001 & 3803.237 & 3803.242 \\
$R_p/R_*$ & $\ldots$ & $\mathcal{U}(0; 1)$ & 0.08954 & 0.08962 & 0.00268 & 0.08478 & 0.09488 \\
$b$ & $\ldots$ & $\mathcal{U}(0; 1.1)$ & 0.2849 & 0.2941 & 0.1709 & 0.02122 & 0.6119 \\
$q_{1,\mathrm{TESS}}$ & $\ldots$ & $\mathcal{U}(0; 1)$ & 0.289 & 0.323 & 0.204 & 0.0334 & 0.778 \\
$q_{2,\mathrm{TESS}}$ & $\ldots$ & $\mathcal{U}(0; 1)$ & 0.415 & 0.439 & 0.262 & 0.0312 & 0.937 \\
$\rho_*$ & g cm$^{-3}$ & $\mathcal{J}(0.1; 50)$ & 1.1045 & 1.0563 & 0.1887 & 0.6409 & 1.309 \\
$\sigma_{\mathrm{GP,TESS}}$ & $\ldots$ & $\mathcal{J}(1e-06; 1)$ & 0.000127 & 0.000132 & 2.82e-05 & 9.27e-05 & 0.000194 \\
$C_{\mathrm{GP,TESS}}$ & $\ldots$ & $\mathcal{U}(0; 1)$ & $\ldots$ & $\ldots$ & $\ldots$ & $\ldots$ & $<0.129^{\rm d}$ \\
$L_{\mathrm{GP,TESS}}$ & day & $\mathcal{J}(0.5; 100)$ & 2.05 & 2.13 & 0.483 & 1.45 & 3.22 \\
$P_{\mathrm{rot,TESS}}$ & day & $\mathcal{N}(1.83; 0.5)$ & 1.461 & 1.463 & 0.05354 & 1.366 & 1.567 \\
\hline
\multicolumn{8}{l}{\textit{Sampled: nuisance}} \\
$\Delta f_{\mathrm{SWOPE}}$ & $\ldots$ & $\mathcal{N}(0; 0.1)$ & -0.003469 & -0.003474 & 0.000475 & -0.00436 & -0.002594 \\
$\sigma_{w,\mathrm{SWOPE}}$ & ppm & $\mathcal{J}(0.1; 10000)$ & 5.03e+03 & 5.04e+03 & 370 & 4.37e+03 & 5.77e+03 \\
$\theta_{0,\mathrm{SWOPE}}$ & day$^{{-1}}$ & $\mathcal{U}(-1; 1)$ & 0.0439 & 0.044 & 0.00851 & 0.0284 & 0.0604 \\
$\theta_{1,\mathrm{SWOPE}}$ & day$^{{-2}}$ & $\mathcal{U}(-1; 1)$ & $\ldots$ & $\ldots$ & $\ldots$ & $\ldots$ & $<-0.826^{\rm d}$ \\
$\Delta f_{\mathrm{SSO}}$ & $\ldots$ & $\mathcal{N}(0; 0.1)$ & -0.004908 & -0.004915 & 0.000483 & -0.005829 & -0.004016 \\
$\sigma_{w,\mathrm{SSO}}$ & ppm & $\mathcal{J}(0.1; 10000)$ & 2.68e+03 & 2.69e+03 & 199 & 2.33e+03 & 3.08e+03 \\
$\theta_{0,\mathrm{SSO}}$ & day$^{{-1}}$ & $\mathcal{U}(-1; 1)$ & 0.00635 & 0.00639 & 0.00465 & -0.00224 & 0.0152 \\
$\theta_{1,\mathrm{SSO}}$ & day$^{{-2}}$ & $\mathcal{U}(-1; 1)$ & 0.186 & 0.185 & 0.0704 & 0.0503 & 0.318 \\
$\Delta f_{\mathrm{CTIOz}}$ & $\ldots$ & $\mathcal{N}(0; 0.1)$ & -0.005684 & -0.005686 & 0.0005323 & -0.006663 & -0.004687 \\
$\sigma_{w,\mathrm{CTIOz}}$ & ppm & $\mathcal{J}(0.1; 10000)$ & 2.2e+03 & 2.2e+03 & 174 & 1.89e+03 & 2.54e+03 \\
$\theta_{0,\mathrm{CTIOz}}$ & day$^{{-1}}$ & $\mathcal{U}(-1; 1)$ & -0.0834 & -0.0834 & 0.00249 & -0.0882 & -0.0788 \\
$\theta_{1,\mathrm{CTIOz}}$ & day$^{{-2}}$ & $\mathcal{U}(-1; 1)$ & 0.106 & 0.105 & 0.0556 & -0.00256 & 0.208 \\
$\Delta f_{\mathrm{CTIOg}}$ & $\ldots$ & $\mathcal{N}(0; 0.1)$ & -0.002847 & -0.002847 & 0.001284 & -0.005275 & -0.0003776 \\
$\sigma_{w,\mathrm{CTIOg}}$ & ppm & $\mathcal{J}(0.1; 10000)$ & 11.8 & 331 & 819 & 0.143 & 2.89e+03 \\
$\theta_{0,\mathrm{CTIOg}}$ & day$^{{-1}}$ & $\mathcal{U}(-1; 1)$ & 0.0132 & 0.0131 & 0.0113 & -0.00833 & 0.0341 \\
$\theta_{1,\mathrm{CTIOg}}$ & day$^{{-2}}$ & $\mathcal{U}(-1; 1)$ & 0.289 & 0.288 & 0.186 & -0.0664 & 0.639 \\
$\Delta f_{\mathrm{mcdg}}$ & $\ldots$ & $\mathcal{N}(0; 0.1)$ & -0.006072 & -0.00608 & 0.001063 & -0.008092 & -0.004068 \\
$\sigma_{w,\mathrm{mcdg}}$ & ppm & $\mathcal{J}(0.1; 10000)$ & 215 & 1.07e+03 & 1.34e+03 & 0.194 & 3.98e+03 \\
$\theta_{0,\mathrm{mcdg}}$ & day$^{{-1}}$ & $\mathcal{U}(-1; 1)$ & -0.0242 & -0.0242 & 0.0167 & -0.0557 & 0.00746 \\
$\theta_{1,\mathrm{mcdg}}$ & day$^{{-2}}$ & $\mathcal{U}(-1; 1)$ & -0.00892 & -0.00225 & 0.348 & -0.637 & 0.686 \\
$\Delta f_{\mathrm{mcdi}}$ & $\ldots$ & $\mathcal{N}(0; 0.1)$ & -0.008185 & -0.008193 & 0.000578 & -0.009307 & -0.007145 \\
$\sigma_{w,\mathrm{mcdi}}$ & ppm & $\mathcal{J}(0.1; 10000)$ & 2.76e+03 & 2.76e+03 & 227 & 2.34e+03 & 3.21e+03 \\
$\theta_{0,\mathrm{mcdi}}$ & day$^{{-1}}$ & $\mathcal{U}(-1; 1)$ & -0.0468 & -0.0468 & 0.00748 & -0.0609 & -0.0327 \\
$\theta_{1,\mathrm{mcdi}}$ & day$^{{-2}}$ & $\mathcal{U}(-1; 1)$ & -0.251 & -0.252 & 0.163 & -0.559 & 0.055 \\
$\Delta f_{\mathrm{ctioaprg}}$ & $\ldots$ & $\mathcal{N}(0; 0.1)$ & -0.001714 & -0.001723 & 0.0006567 & -0.002963 & -0.000497 \\
$\sigma_{w,\mathrm{ctioaprg}}$ & ppm & $\mathcal{J}(0.1; 10000)$ & 9.29 & 133 & 306 & 0.139 & 1.01e+03 \\
$\theta_{0,\mathrm{ctioaprg}}$ & day$^{{-1}}$ & $\mathcal{U}(-1; 1)$ & -0.0505 & -0.0504 & 0.0126 & -0.0739 & -0.0261 \\
$\theta_{1,\mathrm{ctioaprg}}$ & day$^{{-2}}$ & $\mathcal{U}(-1; 1)$ & -0.0488 & -0.0467 & 0.35 & -0.71 & 0.625 \\
$\Delta f_{\mathrm{ctioapri}}$ & $\ldots$ & $\mathcal{N}(0; 0.1)$ & -0.0009538 & -0.0009591 & 0.0002863 & -0.001509 & -0.0004391 \\
$\sigma_{w,\mathrm{ctioapri}}$ & ppm & $\mathcal{J}(0.1; 10000)$ & 1.21e+03 & 1.21e+03 & 220 & 807 & 1.63e+03 \\
$\theta_{0,\mathrm{ctioapri}}$ & day$^{{-1}}$ & $\mathcal{U}(-1; 1)$ & -0.0337 & -0.0336 & 0.00621 & -0.045 & -0.0216 \\
$\theta_{1,\mathrm{ctioapri}}$ & day$^{{-2}}$ & $\mathcal{U}(-1; 1)$ & -0.157 & -0.156 & 0.143 & -0.421 & 0.113 \\
$q_{1,\mathrm{SWOPE}}$ & $\ldots$ & $\mathcal{U}(0; 1)$ & 0.446 & 0.460 & 0.269 & 0.0307 & 0.946 \\
$q_{2,\mathrm{SWOPE}}$ & $\ldots$ & $\mathcal{U}(0; 1)$ & 0.297 & 0.345 & 0.248 & 0.017 & 0.886 \\
$q_{1,\mathrm{CTIOz}}$ & $\ldots$ & $\mathcal{U}(0; 1)$ & 0.351 & 0.376 & 0.195 & 0.0796 & 0.804 \\
$q_{2,\mathrm{CTIOz}}$ & $\ldots$ & $\mathcal{U}(0; 1)$ & 0.279 & 0.325 & 0.24 & 0.0152 & 0.862 \\
$q_{1,\mathrm{CTIOg_mcdg_ctioaprg}}$ & $\ldots$ & $\mathcal{U}(0; 1)$ & 0.768 & 0.748 & 0.164 & 0.408 & 0.984 \\
$q_{2,\mathrm{CTIOg_mcdg_ctioaprg}}$ & $\ldots$ & $\mathcal{U}(0; 1)$ & 0.789 & 0.755 & 0.179 & 0.351 & 0.988 \\
$q_{1,\mathrm{SSO_mcdi_ctioapri}}$ & $\ldots$ & $\mathcal{U}(0; 1)$ & $\ldots$ & $\ldots$ & $\ldots$ & $\ldots$ & $<0.59^{\rm d}$ \\
$q_{2,\mathrm{SSO_mcdi_ctioapri}}$ & $\ldots$ & $\mathcal{U}(0; 1)$ & 0.239 & 0.305 & 0.25 & 0.0105 & 0.888 \\
\hline\hline
\end{tabular}

\vspace{0.5em}

\begin{minipage}{\textwidth}
\footnotesize

\textit{Notes.} $\mathcal{N}$ denotes a normal distribution, $\mathcal{U}$ a uniform distribution, and $\mathcal{J}$ a log-uniform (Jeffreys) distribution. 

Parameters marked with superscript $^{\rm d}$ are reported as 97th-percentile upper limits because the posterior is pinned against the lower edge of its prior (i.e., there is no evidence for excess white noise beyond that already accounted for by the GP model and/or the formal photometric uncertainties), making a symmetric median $\pm$ uncertainty inappropriate.
\end{minipage}

\end{table*}

%% file: table5b.tex
\begin{table}
    \centering
    \caption{Derived Planetary Parameters}
    \label{tab:derived}
    \small
    \setlength{\tabcolsep}{3pt}
    \begin{tabular}{lcccccc}
    \hline\hline
    Parameter & Unit & Median & Mean & Std. Dev. & 3\% & 97\% \\
    \hline
    $a/R_*$      & $\ldots$     & 10.8   & 10.6 & 0.677 & 9.008   & 11.43 \\
    $a$         & A.U. & 0.0367 & 0.0363 & 0.0026 & 0.0306 & 0.0403 \\
    $T_{\rm eq}$     & K             & 684.25   & 698.42 & 26.09 & 652.07 & 750.05 \\ 
    Insolation & $S_\oplus$   & 36.36    & 37.42  & 7.01  & 27.13  & 53.64 \\ 
    $R_p$         & $R_\oplus$   & 7.20   & 7.21 & 0.32 & 6.62    & 7.83 \\
    Depth & ppt  & 8.017 & 8.038 & 0.481 & 7.187 & 9.002 \\  
    $T_{14}$      & hr           & 3.468  & 3.473 & 0.06746 & 3.358   & 3.613 \\
    $T_{23}$      & hr           & 2.83   & 2.82 & 0.0768 & 2.64    & 2.93 \\
    \hline\hline
    \end{tabular}
    \end{table}